\documentclass[twocolumn,aps,prd,showpacs,nofootinbib,floatfix,eqsecnum]{revtex4}
\usepackage{amsmath,latexsym,graphicx,multirow,psfrag}
\begin{document}
\title{Measuring coalescing massive binary black holes with
gravitational waves:\\ The impact of spin-induced precession}
\author{Ryan N.\ Lang and Scott A.\ Hughes}
\affiliation{Department of Physics and MIT Kavli Institute, MIT, 77
Massachusetts Ave., Cambridge, MA 02139}

\begin{abstract}
The coalescence of massive black holes generates gravitational waves
(GWs) that will be measurable by space-based detectors such as LISA to
large redshifts.  The spins of a binary's black holes have an
important impact on its waveform.  Specifically, geodetic and
gravitomagnetic effects cause the spins to precess; this precession
then modulates the waveform, adding periodic structure which encodes
useful information about the binary's members.  Following pioneering
work by Vecchio, we examine the impact upon GW measurements of
including these precession-induced modulations in the waveform model.
We find that the additional periodicity due to spin precession breaks
degeneracies among certain parameters, greatly improving the accuracy
with which they may be measured.  In particular, mass measurements are
improved tremendously, by one to several orders of magnitude.
Localization of the source on the sky is also improved, though not as
much --- low redshift systems can be localized to an ellipse which is
roughly $\mbox{a few} \times 10$ arcminutes in the long direction and
a factor of $2-4$ smaller in the short direction.  Though not a
drastic improvement relative to analyses which neglect spin
precession, even modest gains in source localization will greatly
facilitate searches for electromagnetic counterparts to GW events.
Determination of distance to the source is likewise improved: We find
that relative error in measured luminosity distance is commonly $\sim
0.2\%-0.7\%$ at $z \sim 1$.  Finally, with the inclusion of precession,
we find that the magnitude of the spins themselves can typically be
determined for low redshift systems with an accuracy of about $0.1\%-10
\%$, depending on the spin value, allowing accurate surveys of mass
and spin evolution over cosmic time.
\end{abstract}
\maketitle

\section{Introduction}
\label{sec:intro}

\subsection{Background to this analysis}

Observations have now demonstrated that massive black holes are
ubiquitous in the local universe. It appears that all galaxies with
central bulges contain black holes whose masses are strongly
correlated with the properties of the bulge \cite{fm00, g00}.
Hierarchical structure formation teaches us that these galaxies
assembled over cosmic history through the repeated coalescence of the
dark matter halos in which they reside {\cite{hierarchy}}.  Taken
together, these suggest that coalescences of massive black holes
should be relatively frequent events, especially at high redshift when
halo coalescences were common \cite{vmh03}.

Massive black hole coalescences are extremely strong gravitational
wave (GW) sources.  In the relevant mass band --- thousands to
millions of solar masses --- the GWs these binaries generate are at
low frequency ($f \sim 10^{-4.5} - 10^{-1}$ Hz) where ground-based GW
antennae have poor sensitivity due to geophysical and other
terrestrial noise sources.  Measuring GWs from massive black holes
requires going into the quiet environment of space.  LISA, the Laser
Interferometer Space Antenna, is being designed as a joint NASA-ESA
mission to measure GWs in this frequency band; cosmological massive
black hole coalescences are among its highest priority targets.  By
measuring these GWs, one can infer the properties of the source that
generated the waves.  Some particularly important and interesting
properties are the masses of the binary's members, their spins, the
binary's location on the sky, and its distance from the solar system
barycenter.  Measuring a population of coalescence events could thus
provide a wealth of data on the cosmological distribution and
evolution of black hole masses and spins.

Most analyses of how well binary black hole parameters can be
determined by LISA measurements have ignored the impact of
spin-induced precession \cite{c98,h02,bbw05}.  Under such an
assumption, subsets of parameters can be highly correlated with each
other, increasing the errors in parameter estimation.  One such subset
comprises the binary's ``chirp mass'' $\mathcal{M}$, its reduced mass
$\mu$, and the spin parameters $\beta$ and $\sigma$ (which are written
out explicitly in Sec. \ref{sec:intrinsic}).  These four parameters
influence the GW phase $\Phi$.  As discussed in \cite{cf94,pw95}, the
correlation coefficient between $\mu$ and $\beta$ is nearly 1.  It is
thus difficult to ``detangle'' these parameters from one another in a
measurement.

Another such subset consists of a binary's sky position, orientation,
and distance.  To see why these parameters are strongly correlated,
consider the form of the two polarizations of the strongest quadrupole
harmonic of the gravitational waveform:

\begin{align}
h_+(t) &= 2 \frac{{\mathcal M}^{5/3}(\pi f)^{2/3}}{D_L}(1+\cos^2\iota)
\cos\Phi(t) \, ,
\label{eq:hplusintro}\\
h_{\times}(t) &= -4 \frac{{\mathcal M}^{5/3}(\pi f)^{2/3}}{D_L}\cos
\iota \sin\Phi(t) \, .
\label{eq:hcrossintro}
\end{align}
(We work in units with $G = 1 = c$; a convenient conversion factor in
this system is $10^6\,M_\odot = 4.92\,{\rm seconds}$.)  The quantity
$\iota$ is the binary's inclination relative to the line of sight:
$\cos\iota \equiv \mathbf{\hat{L}} \cdot \mathbf{\hat{n}}$, where
$\mathbf{\hat{L}}$, the direction of the binary's orbital angular
momentum, defines its orientation and $\mathbf{\hat{n}}$ is the
direction from observer to source.  The quantity $D_L$ is the
luminosity distance to the source, and $f(t) \equiv (1/2\pi)
d\Phi/dt$.

One does not measure the polarizations $h_+$ and $h_\times$ directly;
rather, one measures a sum $h_M(t)$ in which the two polarizations are
weighted by antenna response functions as follows:
\begin{equation}
h_M(t) = F_+(\theta_N,\phi_N,\psi_N) h_+(t) +
F_\times(\theta_N,\phi_N,\psi_N) h_\times(t) \;.
\label{eq:hM_schem}
\end{equation}
(This equation should be taken as schematic; see \ref{sec:extrinsic}
for a more detailed and definitive description.)  The angles
$\theta_N$ and $\phi_N$ denote the location of the source on the sky
in some appropriate coordinate system.  The angle $\psi_N$, known as
the ``polarization angle,'' fixes the orientation of the component of
$\mathbf{\hat{L}}$ perpendicular to the line of sight.  (In other
words, $\mathbf{\hat{L}}$ is fixed by $\iota$ and $\psi_N$.)

Measuring the phase determines chirp mass with high accuracy; the
fractional error in ${\cal M}$ is often $\sim 10^{-3} - 10^{-4}$.  As
far as amplitude is concerned, the chirp mass can be regarded as
measured exactly.  What remains is to determine, from the measured
amplitude and the known ${\cal M}$, the angles $\theta_N$, $\phi_N$,
$\psi_N$, $\iota$, and the distance $D_L$.

As Eqs.\ \eqref{eq:hplusintro}, \eqref{eq:hcrossintro}, and
\eqref{eq:hM_schem} illustrate, these five parameters are strongly
correlated.  The motion of LISA around the sun\footnote{LISA is being
designed as a constellation of three spacecraft whose centroid orbits
the sun with a period of one year; see {\tt http://lisa.nasa.gov} for
further details.} breaks these degeneracies to some extent --- the
angles $\theta_N$ and $\phi_N$ appearing in Eq.\ (\ref{eq:hM_schem})
can be regarded as best defined in a coordinate system tied to LISA.
As the antenna orbits the sun, these angles become effectively time
dependent.  The one-year periodicity imposed by this motion makes it
possible to detangle these parameters.  Analyses typically find that
the position of a merger event at $z \sim 1$ can be determined, on
average, to an ellipse which is $1.5-2$ degrees across in the long
direction and $1.5-2$ times smaller in the short direction\footnote{It
is worth bearing in mind that the full moon subtends an angle of about
30 arcminutes.} {\cite{c98,bbw05,hh05}}.  The distance to such a
binary can be determined to $1\%-2\%$ accuracy on average (less in some
exceptional cases) \cite{h02,bbw05,hh05}.

\subsection{Black hole spin and spin precession}

The preceding discussion ignores an important piece of relativistic
physics: the precession of each binary member's spin vector due to its
interaction with the spacetime in which it moves.  In general
relativity, the spacetime of an isolated object can be regarded as
having an ``electric piece,'' arising from the object's mass and mass
distribution, and a ``magnetic piece,'' arising from the object's mass
currents and their distribution\footnote{This analogy is most apt in
the weak field.  In that limit, one can recast the Einstein field
equations of general relativity into a form quite similar to Maxwell's
equations; see {\cite{membrane}} for detailed discussion.  Though the
analogy does not fit quite so well in strong field regions, it remains
accurate enough to be useful.}.  Spin precession consists of a
geodetic term, arising from the parallel transport of the spin vector
in the gravitoelectric field of the other hole, and Lense-Thirring
terms, caused by the gravitomagnetic field of the other hole.  The
basic physics of gravitomagnetic precession can be simply understood
by analogy with a similar (and closely related) electromagnetic
phenomenon --- the precession of a magnetic dipole
{\mbox{\boldmath$\mu$}} immersed in an external magnetic field ${\bf
B}$.  An object's spin angular momentum ${\bf S}$ can be regarded as a
gravitational ``magnetic dipole.''  When immersed in a
``gravitomagnetic field,'' one finds that ${\bf S}$ feels a torque,
just as a magnetic dipole {\mbox{\boldmath$\mu$}} experiences a torque
when immersed in magnetic field ${\bf B}$.  In a binary black hole
system, the gravitomagnetic field arises from the binary's orbital
motion and the spins of its members.  Precession thus includes both
spin-orbit (geodetic and orbital gravitomagnetic) and spin-spin
effects \cite{th85}.  (The major goal of the ``Gravity Probe B''
experiment is to measure the effects of geodetic and spin-spin
Lense-Thirring precession upon a gyroscope in low Earth orbit
{\cite{gpb}}.)

\begin{figure}[htb]
\psfragscanon
\psfrag{xlabel}{$t$ (s)}
\psfrag{ylabel}{$A_{\mathrm{pol, I}}(t)$}
\includegraphics[scale=0.54]{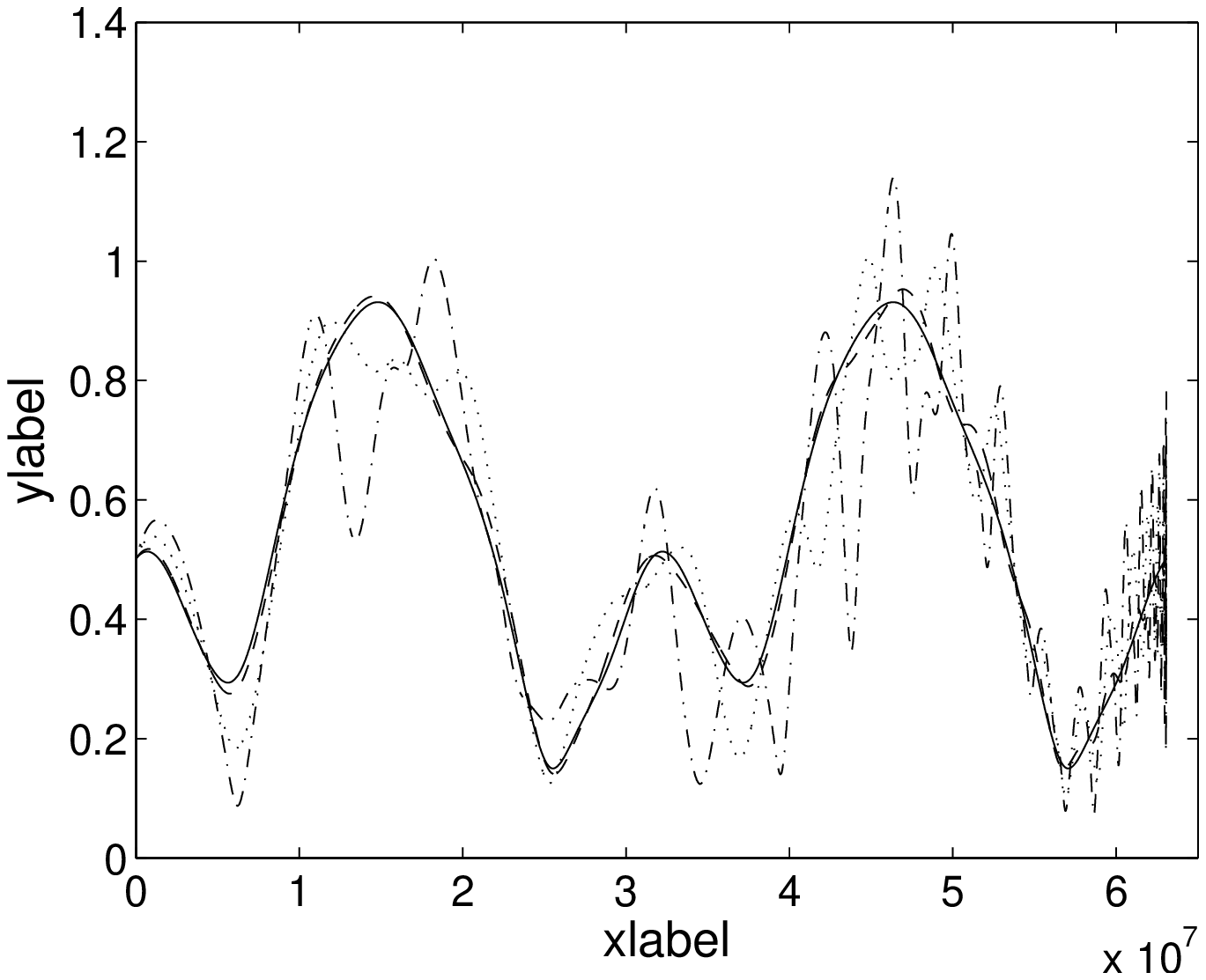}
\includegraphics[scale=0.54]{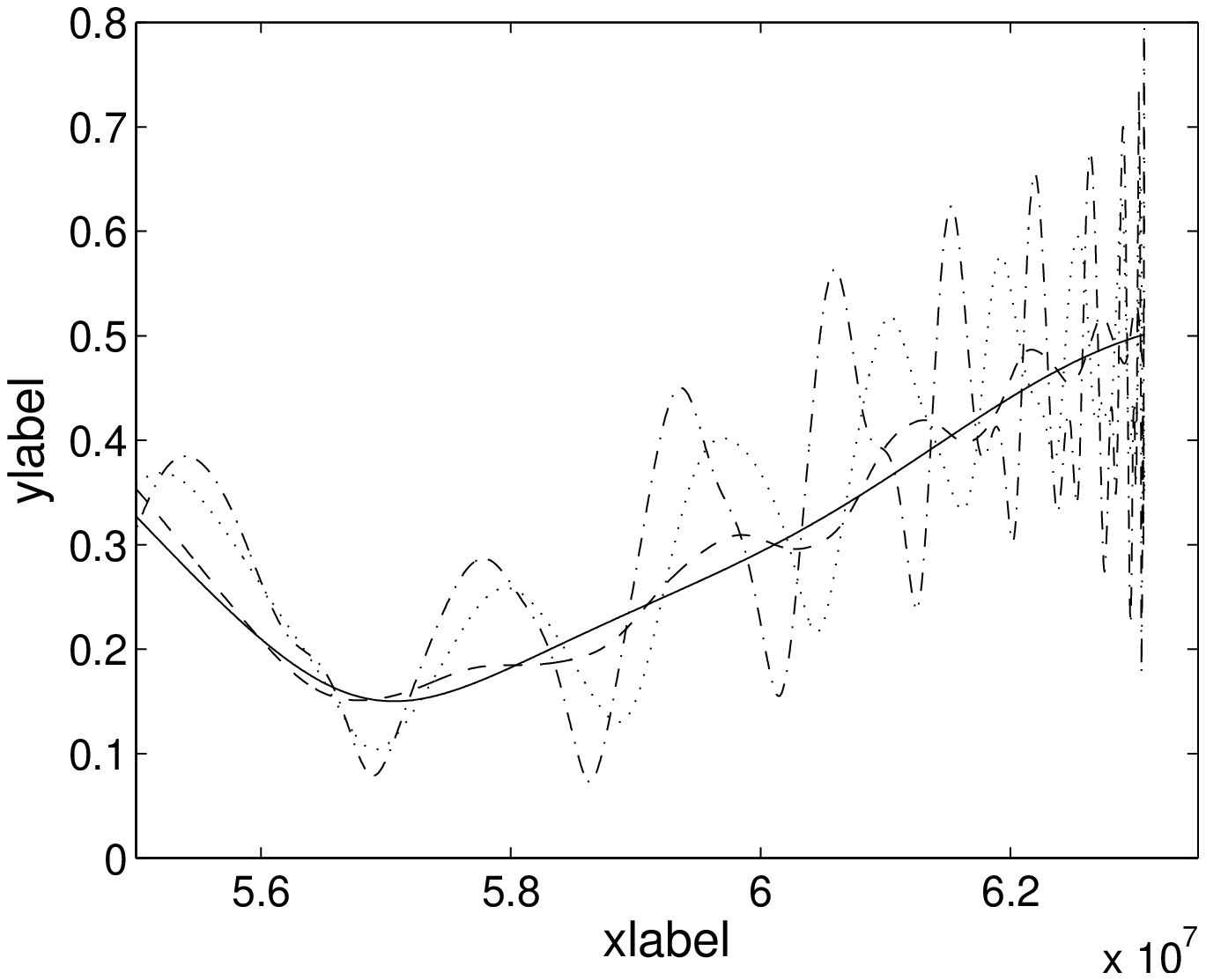}
\caption{These figures depict the ``polarization amplitude''
$A_{\mathrm{pol}}(t)$ of the signal measured in detector I as a
function of time.  The curves are as follows: solid line, $\chi_1 =
\chi_2 = 0$; dashed line (nearly overlapping the solid line), $\chi_1
= \chi_2 = 0.1$; dotted line, $\chi_1 = \chi_2 = 0.5$; and dot-dashed
line, $\chi_1 = \chi_2 = 0.9$.  ($\chi = S/m^2$ is the dimensionless
spin parameter.)  The top figure covers the last two years of
inspiral.  The spinless curve has periodicity of one year,
corresponding to the motion of LISA around the sun.  Notice that as
spin is introduced, the curves become more strongly modulated, with
the number of additional oscillations growing as the spin is
increased.  By tracking these spin-precession-induced modulations, it
becomes possible to better measure parameters like mass and sky
position and measure spin for the first time.  The bottom figure shows
a close-up of the final months of inspiral.}
\label{fig:Apcomparison}
\end{figure}

As the spins precess, they do so in such a way that the {\it total}
angular momentum ${\mathbf{J}} = {\mathbf{L}} + {\mathbf{S}}_1 +
{\mathbf{S}}_2$ is held constant; the orbital angular momentum
${\mathbf{L}}$ precesses to compensate for changes in ${\mathbf{S}}_1$
and ${\mathbf{S}}_2$.  As a consequence, the inclination angle $\iota$
and polarization angle $\psi_N$ become time varying (as do certain
other quantities appearing in the GW phase function $\Phi$).  Figure
\ref{fig:Apcomparison} shows the so-called ``polarization amplitude,''
defined in Sec.\ \ref{sec:extrinsic}, of the waveform measured by a
particular detector.  Without precession, this quantity is modulated
by the orbital motion of LISA, helping to provide some information
about the binary's sky position.  The polarization amplitude also
depends on the angles $\iota$ and $\psi_N$, so it undergoes additional
modulation when precession is included.  Such precession-imposed time
variations quite thoroughly break many of the degeneracies which have
been found to limit parameter measurement accuracy in earlier
analyses.

It is without a doubt that black holes in nature spin.  Observations
are not yet precise enough to indicate the value of typical black hole
spins; the evidence to date does, however, seem to indicate that
fairly rapid rotation is common.  For example, the existence of jets
from active systems seems to require non-negligible black hole spin
--- jets appear to be ``launched'' by the shearing of magnetic field
lines (supported by the highly conductive, ionized material accreting
onto the black hole) by the differential rotation of spacetime around
a rotating black hole {\cite{bz77,m03}}.  Also, observations of highly
distorted iron K-$\alpha$ lines --- a very sharp flourescence feature
in the rest frame of the emitting iron ions --- indicate that this
emission is coming from very deep within a gravitational potential (at
radii less than the Schwarzschild innermost radius $6M$) and is
smeared by near luminal relativistic speeds to boot \cite{rn03}.
Though perhaps influenced somewhat by selection effects\footnote{The
systems for which we have constraints on spin are systems which are
actively accreting, and, {\it ipso facto}, those which are most likely
to be rapidly spinning \cite{hb03}.}, these pieces of evidence are
strong hints that the black holes which will form the binaries we hope
to measure will be strongly influenced by spin.

The only limit in which spin precession can be neglected is that in
which the spins of the binary's members are exactly parallel (or
antiparallel) to one another and to the orbital angular momentum
$\mathbf{L}$.  Since the target binaries of this analysis are created
by galactic merger processes, their members will almost certainly have
no preferred alignment --- random spin and orbit orientation is
expected to be the rule.  (This expectation is borne out by work
{\cite{spck02}} showing that jets in active galaxies are oriented
randomly with respect to the disks of their host galaxies.)  Taking
into account spin precession is thus of paramount importance for GW
observations of merging black hole systems.

A great deal of work has gone into developing families of model
waveforms (``templates'') sufficiently robust to {\it detect} GWs from
spinning and precessing binaries, at least in the context of
measurements by ground-based detectors
\cite{bcv03,bcv041,bcv042,bcv05,gkv03,gk03,gikb04}.  The key issue in
this case is that the various modulations on the waveform imposed by
the binary's precession smear its power over a wider spectral range,
making it much more difficult to detect at the (relatively) low
signal-to-noise ratios (SNRs) expected for ground-based observations.
Not as much work has gone into the complementary problem of {\it
measuring} these waves --- examining the impact precession has upon
the precision with which binary properties may be inferred from the
waves.  To date, the most complete and important analysis of this type
is that of Vecchio {\cite{v04}}.  Vecchio focuses (for simplicity) on
equal mass binaries and only includes the leading ``spin-orbit''
precession term.  This limit is particularly nice as a first analysis
of this problem, since it can be treated (largely) analytically (cf.\
discussion in Sec.\ III B of Ref.\ {\cite{v04}}).

Vecchio's work largely confirms the intuitive expectation discussed
above --- the precision with which masses are measured is
substantially improved; in particular, the reduced mass of the system
can be measured with several orders of magnitude more accuracy.
Parameters such as the sky location of the binary and the luminosity
distance are also measured more accurately, but only by a factor of 2
-- 10.

\subsection{This analysis}

Our goal here is to update Vecchio's pioneering analysis by taking the
precession equations and the wave phase to the next higher order and
by performing a broader parameter survey (including the impact of mass
ratio).  By taking the precession equations to higher order, we
include ``spin-spin'' effects --- precessional effects due to one
black hole's spin interacting with gravitomagnetic fields from the
other hole's spin.  By taking the wave phase to higher order, we
include, among other terms, a time-dependent spin-spin interaction.
Finally, when the mass ratio differs from 1, the geodetic spin-orbit
term causes the two spins to precess at different rates, even without
the spin-spin corrections.
  
Including these effects means that we {\it cannot} model the
precession with a simple, analytic rule --- we are forced to integrate
the equations of precession numerically as inspiral proceeds,
incurring a significant performance cost.  Fortunately, the basic
``engine'' on which this code is based {\cite{h02}} runs extremely
fast, thanks largely to the use of spectral integrators (which, in
turn, is thanks to a suggestion by E.\ Berti {\cite{bbw05}}), so total
run time remains reasonable.

The cost in efficiency due to the inclusion of higher-order effects is
offset by the more complete description of the signal they provide.
An important consequence is that it now becomes possible from GW
measurements to determine the spin of each member of the binary.  With
Vecchio's approximations, only three components of the black holes'
vector spins can be determined --- enough to constrain, but not
determine, their spin magnitudes.  Our more general approach allows us
to measure all six vector spin components.  To our knowledge, this is
the first analysis indicating how well spin can be measured from
merging comparable mass binary systems.  (As Barack and Cutler have
shown {\cite{bc04}}, spin is very well determined by measurements of
GWs from {\it extreme mass ratio} binaries --- those in which the
system's mass ratio $m_2/m_1 \lesssim 10^{-4}$ or so.)

Our error estimates are computed using the maximum likelihood
formalism first introduced in the context of GW measurements by Finn
{\cite{f92}}.  A potential worry is that we are using a Gaussian
approximation to the likelihood function.  This approximation is very
convenient since it allows us to directly compute a Fisher information
matrix.  Its inverse is the covariance matrix, which directly encodes
the estimated 1-$\sigma$ errors in measured parameters, as well as
correlations among different parameters.  The Gaussian approximation
is known to be accurate when the SNR is ``high enough''
\cite{f92,cf94}.

Unfortunately, it is not particularly obvious what ``high enough''
really means.  In our case, we are estimating measurement errors on 15
parameters\footnote{Two masses; 2 angles specifying the initial
orientation of the binary's orbit; 4 angles specifying the initial
orientation of the spins; 2 spin magnitudes; the time at which
coalescence occurs; the phase at coalescence; 2 angles specifying the
binary's position on the sky; and the distance to the binary.}  --- a
rather fearsome number to fit.  The Gaussian approximation almost
certainly {\it underestimates} measurement error, since it assumes the
likelihood function is completely determined by its curvature in the
vicinity of a maximum, missing the possibility of a long tail to large
error.  We thus fear that our estimates are likely to be optimistic,
especially for events with relatively small SNR.  It would be quite
salubrious to ``spot check'' a few cases by directly computing the
likelihood function in a few important corners of parameter space and
comparing to the Gaussian predictions.  This would both quantify the
degree to which our calculations are too optimistic and help to
determine how large SNR must be for this approximation to be reliable.

In addition to concerns about the Gaussian approximation, it must be
noted that the waveform family we use for our analysis is somewhat
limited.  We use a post-Newtonian description of the GWs from these
binaries.  Since our analysis requires us to follow these binaries
deep into the strong field where the usual post-Newtonian expansion is
likely to be somewhat unreliable, it is likely that we are introducing
some systematic error.  In particular, the equations of spin
precession that we are using are only given to the leading order
needed to see spin-orbit and spin-spin precession effects
\cite{acst94}.  Higher spin-orbit corrections to the equations of
motion and precession have recently been derived {\cite{fbb06}}, as
have their impact on the the waves' phasing {\cite{bbf06}}.  Another
analysis {\cite{pr06}} has worked out higher-order spin-spin
corrections to the post-Newtonian metric, from which it would not be
too difficult to work out equations of motion and precession and then
the modification to the waves' phase.  It would be interesting to see
what effect the higher-order corrections have on these results.

Finally, it should be noted that the frequency domain expression of
the signal which we are using is derived formally using a ``stationary
phase'' approximation.  This approximation is based on the idea that
the binary's orbital frequency is changing ``slowly.''  The orbital
frequency is thus well-defined over ``short'' time scales.
Quantitatively, this amounts to a requirement that the time scale on
which radiation reaction changes the orbital frequency, $T_{\rm
insp}$, be much longer than an orbital period, $T_{\rm orb}$.
Precession introduces a new time scale, $T_{\rm prec}$, the time it
takes for the angular momentum vectors to significantly change their
orientations.  For the stationary phase approximation to be accurate,
we must in addition require $T_{\rm prec} \gg T_{\rm orb}$, a somewhat
more stringent requirement than $T_{\rm insp} \gg T_{\rm orb}$.  No
doubt, a certain amount of error is introduced due to the breakdown of
this condition late in the inspiral.

Thus, the results which we present here should be taken as {\it
indicative} of how well LISA is likely to be able to measure the
parameters of massive black hole binaries, but cannot be considered
definitive.  We are confident however that the {\it improvement} in
measurement accuracy obtained by taking spin precession into account
is robust.  Specifically, we see that errors in masses are reduced
dramatically, from one to several orders of magnitude.  Errors in sky
position and distance are also reduced, but by a smaller factor.  Such
improvement may nonetheless critically improve the ability of LISA to
interface with electromagnetic observatories
{\cite{an02,mp05,kfhm06}}.  Finally, the added information in the
precession signal allows us to measure the spins of the holes.  These
improvements due to precession will certainly survive and play an
important role even in an analysis which addresses the caveats we list
above.

\subsection{Organization of this paper}

The remainder of the paper is organized as follows.  In Sec.\
\ref{sec:inspiral}, we discuss the gravitational waveform generated by
binary black hole coalescence, focusing upon the slow, adiabatic
inspiral.  Section \ref{sec:intrinsic} describes the ``intrinsic''
waveform produced by the motion of the orbiting black holes as given
in the ``restricted post-Newtonian expansion'' of general relativity.
Section {\ref{sec:precession}} then describes the post-Newtonian
precession equations which we use to model the evolution of the spins
of a binary's members, as well as how those precessions influence the
waveform.  Finally, in Sec.\ {\ref{sec:extrinsic}} we describe
``extrinsic'' effects which enter the measured waveform through its
measurement by the LISA constellation.

In Sec.\ {\ref{sec:meas}} we summarize our parameter estimation
formalism; this section will be largely review to readers familiar
with the literature on GW measurements.  Section
{\ref{sec:meastheory}} first summarizes the maximum likelihood
formalism we use to estimate measurement errors.  In Sec.\
{\ref{sec:lisanoise}}, we then describe the up-to-date model for the
noise which we expect to accompany LISA measurements.

Section {\ref{sec:results}} presents our results.  After describing
some critical procedural issues in the setup of our calculations in
Sec.\ {\ref{sec:procedure}}, we summarize our results for parameters
intrinsic to the binary (particularly masses and spins) in Sec.\
{\ref{sec:resI}} and for extrinsic parameters (particularly sky
position and luminosity distance) in Sec.\ {\ref{sec:resII}}.  In both
cases, we compare, when appropriate, to results from a code which does
not incorporate spin-precession physics.  (This code was originally
developed for the analysis presented in Ref.\ {\cite{h02}}.)  The
general rule of thumb we find is that the accuracy with which masses
can be determined is improved by about one to several orders of
magnitude when precession physics is taken into account.  In addition,
we find that for low redshift ($z \sim 1$) binaries LISA should be
able to determine the spins of the constituent black holes with a
relative precision of $0.1\% - 10\%$, depending (rather strongly) on the
spin value.  Likewise, we find improvement in the measurement accuracy
of extrinsic parameters, though not quite as striking --- half an
order of magnitude improvement in source localization and distance
determination is a good, rough rule of thumb.

An important consequence of these improvements is that LISA should be
able to localize low redshift binaries --- using GW measurements alone
--- to an elliptical ``pixel'' that is perhaps $\mbox{a few} \times
10$ arcminutes across in its widest direction and a factor of $2-4$
smaller along its minor axis.  For higher redshift binaries ($z \sim 3
- 5$), this pixel is several times larger, perhaps a few degrees in
the long direction and tens of arcminutes to a degree or two in the
narrow one.  These results suggest that it should not be too arduous a
task to search for electromagnetic counterparts to a merging binary
black hole's GW signal {\cite{an02,mp05,kfhm06}} --- particularly at
low redshift, these pixel sizes are comparable to the field of view of
planned large scale surveys.

A concluding and summarizing discussion is given in Sec.\
{\ref{sec:disc}}.  Along with summarizing our major results and
findings, we discuss future work which could allow us to
quantitatively assess the consequences of some of the simplifying
assumptions we have made.

At several points in this analysis, we need to convert between a
source's redshift $z$ and luminosity distance $D_L$.  To make this
conversion, we assume a flat cosmology ($\Omega_{\rm total} = 1$) with
contributions from matter ($\Omega_M = 0.25$) and from a cosmological
constant (equation of state parameter $w = -1$, $\Omega_\Lambda =
0.75$).  We also choose a Hubble constant $H_0 = 75 \, \mathrm{km\
s^{-1}\ Mpc^{-1}}$.  These choices are in concordance with the latest
fits presented by the WMAP team in their three-year analysis of the
cosmic microwave background {\cite{setal06}}.  The luminosity distance
as a function of redshift is then given by

\begin{equation}
D_L(z) = \frac{(1+z)c}{H_0}\int_0^z
\frac{dz^\prime}{\sqrt{\Omega_M(1+z')^3 + \Omega_\Lambda}} \, .
\label{eq:Dzrelation}
\end{equation}

\section{Gravitational waves from binary black hole inspiral}
\label{sec:inspiral}

The coalescence of a black hole binary can be divided into three
stages: (1) an adiabatic inspiral, (2) a merger, and (3) a ringdown,
when the resulting black hole settles down to its final state.  In
this paper, we will focus on the inspiral.  Ringdown waves have been
analyzed in other work \cite{e89,f92,h02,dkkfgl04}; the most
comprehensive recent analysis was performed by Berti, Cardoso, and
Will \cite{bcw06}.  The merger waveform, describing the strong field
and (potentially) violent process of the two black holes merging into
a single body, has historically been poorly understood.  Recent
breakthroughs in numerical relativity may soon correct this
\cite{p05,clmz06,bcckv06}.

The inspiral waveform which will be measured by LISA is a combination
of the intrinsic waveform created by the source and extrinsic features
related to its location on the sky and modulation effects caused by
the motion of the detector.  In this section we review the relevant
physics involved in the construction of the waveform.

For sources at cosmological distances, all time scales redshift by a
factor $1+z$.  In the $G = c = 1$ units that we use, all factors of
mass enter as time scales; thus, masses are redshifted by this $1 + z$
factor.  [Likewise, quantities such as spin which have dimension
(time)$^2$ acquire a factor $(1 + z)^2$, etc.]  In the equations
written below, we do not explicitly write out these redshift factors;
they should be taken to be implicit in all our equations.  When
discussing results, we will always quote masses as they would be
measured in the rest frame of the source, with redshift given
separately.

\subsection{Intrinsic waveform}
\label{sec:intrinsic}

We treat the members of our binary as moving on quasicircular orbits;
eccentricity is very rapidly bled away by gravitational radiation
reaction \cite{p64}, so it is expected that these binaries will have
essentially zero eccentricity by the time they enter LISA's frequency
band (at least at the mass ratios we consider in this paper, $1 \le
m_1/m_2 \le 10$).  We use the post-Newtonian formalism, an expansion
in internal gravitational potential $U$ and internal source velocity
$v$, to build our waveforms.  A detailed review of the post-Newtonian
formalism can be found in the article by Blanchet \cite{b02}; the key
pieces which we will use can be found in Refs.\
\cite{b02,kww93,bdiww95,k95,ww96}.

The post-Newtonian equations of motion, taken to second post-Newtonian
(2PN) order, yield the following generalization of Kepler's third law
relating orbital angular frequency $\Omega$ and orbital radius (in
harmonic coordinates) $r$ \cite{bdiww95}:

\begin{equation}
\begin{split}
\Omega^2 &= \frac{M}{r^3}\left[\vphantom{\left(\frac{M}{r}\right)^2}1
- (3 - \eta)\left(\frac{M}{r}\right)\right.\\
& \quad - \sum_{i=1}^2\left(2\frac{m_i^2}{M^2} + 3\eta
\right)\frac{\mathbf{\hat{L}}\cdot
\mathbf{S}_i}{m_i^2}\left(\frac{M}{r}\right)^{3/2}\\
& \quad +\left(6+\frac{41}{4}\eta+\eta^2 \right.\\
& \left. \left. \quad -\frac{3}{2}\frac{\eta}{m_1^2m_2^2}[\mathbf{S}_1
\cdot \mathbf{S}_2 - 3(\mathbf{\hat{L}}\cdot
\mathbf{S}_1)(\mathbf{\hat{L}}\cdot
\mathbf{S}_2)]\right)\left(\frac{M}{r}\right)^2 \right]\, .
\end{split}
\label{eq:PNomega}
\end{equation}
Here $M = m_1 + m_2$ is the total mass of the system, and $\eta =
\mu/M$, where $\mu = m_1m_2/M$ is the reduced mass.
$\mathbf{\hat{L}}$ is the direction of the orbital angular momentum,
and $\mathbf{S}_i$ is the spin angular momentum of black hole $i$.
The magnitude of the spin can be expressed as $S_i = \chi_i m_i^2$,
where $0 \leq \chi_i \leq 1$.  The leading term is the standard result
from Newtonian gravity.  The $O(M/r)$ term is the first post-Newtonian
correction; this is the same physics that, in solar system dynamics,
causes the precession of the perihelion of Mercury.  The
$O((M/r)^{3/2})$ term contains spin-orbit corrections to the equation
of motion.  Finally, the $O((M/r)^2)$ term is a 2PN correction, which
also includes spin-spin terms.  From the equations of motion, the
orbital energy of the binary $E$ can also be computed; see
\cite{bdiww95}.

The waveform that we will use is the ``restricted'' 2PN waveform.
This approximation can be understood by writing the waveform (somewhat
schematically) as \cite{cf94}
\begin{equation}
h(t) = \mathrm{Re}\left(\sum_{x,m}
h^x_m(t)e^{im\Phi_{\mathrm{orb}}(t)}\right) \, ,
\label{eq:restrictedPN}
\end{equation}
where $x$ labels PN order, $m$ is a harmonic index, and $\Phi_{\rm
orb}(t) = \int^t \Omega(t')dt'$ is orbital phase.  In the restricted
post-Newtonian waveform, we throw out all amplitude terms except
$h_2^0$ (the ``Newtonian quadrupole'' term) but compute $\Phi_{\rm
orb}(t)$ to some specified PN order.  The restricted PN approximation
is motivated by the fact that matched filtering --- matching a signal
in noisy data by cross-correlating with a theoretical template --- is
much more sensitive to phase information than to the amplitude.  Since
the $h_2^0$ harmonic contributes most strongly to the waveform over
most of the inspiral, the restricted PN approximation is expected to
capture the most important portion of the inspiral waveform.  It is
worth noting, however, that there is additional information encoded by
the harmonics that we are neglecting.  Especially for the SNRs
expected for typical LISA binary black hole measurements, this
additional information could play an important role in measuring
source characteristics, as pointed out by Hellings and Moore
{\cite{hm03}}.

At any rate, within the restricted PN approximation, the waveform can
be written
\begin{equation}
h_{ij}(t,\mathbf{x}) = -\frac{4 {\mathcal M}^{5/3}(\pi f)^{2/3}}{|\mathbf{x}|} 
\begin{bmatrix}
  \cos \Phi(t) & \sin \Phi(t) & 0 \\
  \sin \Phi(t) & -\cos \Phi(t) & 0 \\
  0 & 0 & 0 \end{bmatrix} \, , 
\label{eq:newth}
\end{equation}
where $|\mathbf{x}|$ is the distance to the binary, ${\mathcal M} =
\mu^{3/5}M^{2/5}$ is the ``chirp mass'' (so called because it largely
determines the rate at which the system's frequency evolves, or
``chirps''), $f = \Omega/\pi = 2f_{\mathrm{orb}}$ is the GW frequency,
and $\Phi(t) = \int^t 2\pi f(t') dt' = 2\Phi_{\mathrm{orb}}$ is the GW
phase.  We have chosen a coordinate system oriented such that the
binary's orbit lies within the $xy$-plane; this tensor will later be
projected onto polarization basis tensors to construct the measured
polarizations $h_+$ and $h_\times$.

The rate at which the frequency changes due to the emission of
gravitational radiation is given by \cite{bdiww95}
\begin{equation}
\begin{split}
\frac{df}{dt} &= \frac{96}{5\pi \mathcal{M}^2}(\pi
\mathcal{M}f)^{11/3}\left[1-\left(\frac{743}{336}+\frac{11}{4}\eta\right)(\pi
Mf)^{2/3} \right. \\
& \quad + (4\pi-\beta)(\pi Mf) \\
& \quad \left. + \left(\frac{34103}{18144} + \frac{13661}{2016}\eta +
\frac{59}{18}\eta^2 + \sigma \right)(\pi Mf)^{4/3}\right] \, .
\end{split}
\label{eq:PNdfdt}
\end{equation}
Notice that the chirp mass ${\cal M}$ dominates the rate of change of
$f$; the reduced mass $\mu$ and parameters $\beta$ and $\sigma$ have
an influence as well.  The parameter $\beta$ describes spin-orbit
interactions and is given by
\begin{equation}
\beta = \frac{1}{12}\sum_{i=1}^2\left[113\left(\frac{m_i}{M}\right)^2
+
75\frac{\mu}{M}\right]\frac{\mathbf{\hat{L}}\cdot\mathbf{S}_i}{m_i^2}
\, .
\label{eq:beta}
\end{equation}
The parameter $\sigma$ describes spin-spin interactions:
\begin{equation}
\sigma = \frac{\mu}{48M(m_1^2m_2^2)}[721(\mathbf{\hat{L}}\cdot
\mathbf{S}_1)(\mathbf{\hat{L}}\cdot
\mathbf{S}_2)-247(\mathbf{S}_1\cdot \mathbf{S}_2)] \, .
\label{eq:sigma}
\end{equation}
Notice that $\beta$ and $\sigma$ depend on the angles between the
binary's angular momentum and the two spins.  In previous analyses
which have neglected precession, $\beta$ and $\sigma$ are constants;
precession makes them time dependent.

Using Eq.\ \eqref{eq:PNdfdt}, we can now integrate to find
\begin{equation}
\begin{split}
t(f) &= t_c -\frac{5}{256}{\mathcal M}(\pi {\mathcal M} f)^{-8/3} \\
& \times \left[1 + \frac{4}{3}\left(\frac{743}{336} +
\frac{11}{4}\eta \right)(\pi Mf)^{2/3}\right.\\
& - \frac{8}{5}(4\pi - \beta)(\pi Mf) \\
& \left. + 2\left(\frac{3058673}{1016064} +
\frac{5429}{1008}\eta + \frac{617}{144}\eta^2 - \sigma\right)(\pi
Mf)^{4/3}\right] \,.
\end{split}
\label{eq:PNtime}
\end{equation}
The parameter $t_c$ formally defines the time at which $f$ diverges
within the post-Newtonian framework.  In reality, we expect finite
size effects to significantly modify the binary's evolution as the
members come into contact.  The system evolves so quickly as the
bodies come together that $t_c$ is nonetheless a useful surrogate for
a ``time of coalescence''.  Finally, the wave phase $\Phi(t) = \int^t
2\pi f(t') dt'$ as a function of $f$ is given by
\begin{equation}
\begin{split}
\Phi(f) &= \Phi[t(f)] = \Phi_c - \frac{1}{16}(\pi {\mathcal M} f)^{-5/3} \\
& \times \left[1 + \frac{5}{3}\left(\frac{743}{336} +
\frac{11}{4}\eta\right)(\pi Mf)^{2/3} \right.\\
& -\frac{5}{2}(4\pi-\beta)(\pi Mf) \\
& \left. + 5\left(\frac{3058673}{1016064} +
\frac{5429}{1008}\eta + \frac{617}{144}\eta^2 - \sigma\right)(\pi
Mf)^{4/3}\right] \, ,
\end{split}
\label{eq:PNphase}
\end{equation}
where $\Phi_c$ is the phase at time $t_c$. The restricted PN waveform
is then constructed by inserting \eqref{eq:PNphase} into \eqref{eq:newth}.

\subsection{Precession equations}
\label{sec:precession}

We next examine the effects of precession on the binary system.  As
discussed in the Introduction, spin-orbit and spin-spin interactions
cause the black hole spins $\mathbf{S}_1$ and $\mathbf{S}_2$ to
precess.  Precession occurs, at leading order\footnote{Several effects
are built in to the precession equations discussed below, leading to
precessions that occur on time scales scaling as $r^{5/2}$ and $r^3$.
Since we integrate these equations numerically, all of these effects
are included in our analysis.  For the purposes of this discussion, we
subsume these effects into the leading-order time scale
$T_{\mathrm{prec}}$ discussed here.}, on a time scale
$T_{\mathrm{prec}} \propto r^{5/2}$ at large separations {\cite{s04}}.
Since this is smaller than the inspiral time scale
$T_{\mathrm{insp}}$, we treat the total angular momentum $\mathbf{J} =
\mathbf{L} + \mathbf{S}_1 + \mathbf{S}_2$ as constant over $T_{\rm
prec}$.  The orbital angular momentum $\mathbf{L}$ must then precess
to compensate for changes in $\mathbf{S}_1$ and $\mathbf{S}_2$.  Since
$T_{\mathrm{prec}}$ is longer than the orbital time scale
$T_{\mathrm{orb}}$, we use an orbit-averaged version of the precession
equations \cite{acst94}:
\begin{align}
  \begin{split}
    \mathbf{\dot{S}}_1 &=
    \frac{1}{r^3}\left[\left(2+\frac{3}{2}\frac{m_2}{m_1}\right)\mu
    \sqrt{Mr} \mathbf{\hat{L}}\right] \times \mathbf{S}_1 \\ & \quad +
    \frac{1}{r^3}\left[\frac{1}{2}\mathbf{S}_2-\frac{3}{2}(\mathbf{S}_2\cdot
    \mathbf{\hat{L}})\mathbf{\hat{L}}\right] \times \mathbf{S}_1 \, ,
  \end{split}
  \label{eq:S1dot}\\
  \begin{split}
    \mathbf{\dot{S}}_2 &= \frac{1}{r^3}\left[\left(2+\frac{3}{2}\frac{m_1}{m_2}\right)\mu \sqrt{Mr} \mathbf{\hat{L}}\right] \times \mathbf{S}_2 \\
    & \quad + \frac{1}{r^3}\left[\frac{1}{2}\mathbf{S}_1-\frac{3}{2}(\mathbf{S}_1\cdot \mathbf{\hat{L}})\mathbf{\hat{L}}\right] \times \mathbf{S}_2 \, ,
  \end{split}
  \label{eq:S2dot}
\end{align}
where\footnote{We use only the lowest-order Newtonian orbital
separation in these equations.  Including more terms would introduce
higher-order effects into the precession.} $r = M^{1/3}/(\pi
f)^{2/3}$.  These equations each have two pieces \cite{th85}.
Consider the equation for $\mathbf{\dot{S}}_1$.  The first piece,
which contains no $\mathbf{S}_2$ dependence, is the spin-orbit term.
This term, which comes in at 1PN order, is due to the geodetic
precession of $\mathbf{S}_1$ as hole 1 orbits in the spacetime
generated by the mass of hole 2, and to the Lense-Thirring precession
of $\mathbf{S}_1$ in the gravitomagnetic field generated by the
orbital motion of hole 2.  The second piece is the spin-spin term,
which enters at 1.5PN order.  This term can be understood as the
Lense-Thirring precession of $\mathbf{S}_1$ in the gravitomagnetic
field generated by the spin of hole 2.  Note that the magnitudes of
the spins do not change at this order; see \cite{acst94} for more
details.  From conservation of total angular momentum on short time
scales, we have

\begin{equation}
  \mathbf{\dot{L}} = -(\mathbf{\dot{S}}_1 + \mathbf{\dot{S}}_2) \, .
  \label{eq:Ldot}
\end{equation}
Over longer time scales, we must also consider the change in total
angular momentum due to the radiation reaction, which is given by
\begin{equation}
\dot{\mathbf{J}} = -
\frac{32}{5}\frac{\mu^2}{r}\left(\frac{M}{r}\right)^{5/2}\mathbf{\hat{L}}
\, 
\end{equation}
to lowest order.

Considering only the spin-orbit terms and taking the limit $S_2 = 0$
or $m_1 = m_2$ leads to a system whose precession can described
analytically; this is the ``simple precession'' limit described in
\cite{acst94}.  Simple precession can be visualized as a rotation of
$\mathbf{L}$ and $\mathbf{S} = \mathbf{S}_1 + \mathbf{S}_2$ around the
total angular momentum $\mathbf{J}$.  (Since inspiral shrinks
$\mathbf{J}$, the precession is actually around a slightly different
direction $\mathbf{J}_0$; see \cite{acst94} for further discussion.)

Since Vecchio restricts his analysis to $m_1 = m_2$ and does not
include the spin-spin interaction, this limit is appropriate for his
work \cite{v04}.  As a consequence, Vecchio takes the quantities
$\mathbf{\hat{L}}\cdot \mathbf{\hat{S}}_i$, $\mathbf{\hat{S}}_1\cdot
\mathbf{\hat{S}}_2$, and $\beta$ to be constant.  (He does not include
the spin-spin term $\sigma$ in the analysis.)  Here, we will study the
impact of the full (albeit orbit-averaged) precession equations,
including spin-spin terms, and include the impact of mass ratio.  An
analytic description is not possible in this case, so we must
integrate these equations numerically.  The behavior is qualitatively
similar to the simple precession case, but with significant
quantitative differences.  For example, $\beta$ now oscillates around
an average value.  For unequal masses (say $m_1/m_2 \gtrsim 2$), the
difference due to precession can be substantial \cite{k95}.  Such
cases are also astrophysically the most interesting --- a mass ratio
of roughly 10 is favored in binary black hole formation scenarios
arising from hierarchical structure formation \cite{shmv04}.

The precession of the orbital plane causes a change in the orbital
phase $\Phi_{\mathrm{orb}}(t)$ {\cite{acst94}}.  Multiplying by a
factor of 2, the change in the wave phase is
\begin{equation} 
\delta_p\Phi(t) = -\int_t^{t_c}\delta_p\dot{\Phi}(t')dt' \, ,
\label{eq:Tomphase}
\end{equation}
where
\begin{equation}
\delta_p\dot{\Phi}(t) = \frac{2\mathbf{\hat{L}}\cdot
\mathbf{\hat{n}}}{1 - (\mathbf{\hat{L}}\cdot
\mathbf{\hat{n}})^2}(\mathbf{\hat{L}}\times \mathbf{\hat{n}})\cdot
\mathbf{\dot{\hat{L}}} \, ,
\label{eq:dTomphase}
\end{equation}
and $\hat{\mathbf{n}}$ is the direction of the binary on the sky.  At
this point, we note that precession has several effects on the
observed waveform.  It changes the orbital phase, even at Newtonian
order, by the amount \eqref{eq:Tomphase}, and it modifies the
functions $\beta$ and $\sigma$ which appear in the post-Newtonian
phase \eqref{eq:PNphase} and time-frequency relation
\eqref{eq:PNtime}.  In the next section, we consider extrinsic effects
on the waveform and find that precession of the orbital plane modifies
them as well.

\subsection{Extrinsic effects}
\label{sec:extrinsic}

We have now constructed the intrinsic GWs emitted by a precessing
binary in the restricted post-Newtonian approximation.  The waveform
measured by LISA will also include extrinsic effects due to the
binary's location on the sky and the motion of the detector.

We can write the wave as a combination of two orthogonal polarizations
propagating in the $-\mathbf{\hat{n}}$ direction.  Define
$\mathbf{\hat{p}}$ and $\mathbf{\hat{q}}$ as axes orthogonal to
$\mathbf{\hat{n}}$, with $\mathbf{\hat{p}} = {\mathbf{\hat{n}}\times
\mathbf{\hat{L}}}/{|\mathbf{\hat{n}}\times \mathbf{\hat{L}}|}$ and
$\mathbf{\hat{q}}$ = $\mathbf{\hat{p}} \times \mathbf{\hat{n}}$.
These are the principal axes for the wave; that is, they are defined
so that the two polarizations are exactly $90^\circ$ out of phase.
The polarization basis tensors for these axes are $H^+_{ij} = p_ip_j -
q_iq_j$ and $H^{\times}_{ij} = p_iq_j + q_ip_j$:
\begin{equation}
h_{ij}(t) = h_+(t)H^+_{ij} + h_{\times}(t)H^{\times}_{ij} \,,
\label{eq:polarizations}
\end{equation}
where
\begin{align}
h_+(t) &= 2 \frac{{\mathcal M}^{5/3}(\pi
f)^{2/3}}{D_L}[1+(\mathbf{\hat{L}}\cdot \mathbf{\hat{n}})^2]
\cos[\Phi(t) + \delta_p \Phi(t)] \, ,
\label{eq:hplus}\\
h_{\times}(t) &= -4 \frac{{\mathcal M}^{5/3}(\pi
f)^{2/3}}{D_L}(\mathbf{\hat{L}}\cdot \mathbf{\hat{n}}) \sin[\Phi(t) +
\delta_p \Phi(t)] \, .
\label{eq:hcross}
\end{align}
These expressions were first discussed in the Introduction (albeit
without the precessional phase correction).  Here $D_L$ is the
luminosity distance to the source.  Notice that the weighting of the
two polarizations depends upon the direction of the angular momentum
vector relative to the sky position.

We now consider the GW as measured by the detector.  All of this
analysis is done using the long wavelength ($\lambda \gg L$, where $L$
is the LISA arm length) approximation introduced by Cutler \cite{c98};
more details can be found there.  This approximation is appropriate
for our purposes since most of the signal accumulates at low
frequencies where the wavelength is in fact greater than the arm
length.  The full LISA response function, including arm-length
effects, is discussed in \cite{cr03,r04}.

LISA consists of three spacecraft arranged in an equilateral triangle,
$5 \times 10^6 \, \mathrm{km}$ apart.  The center of mass of the
configuration orbits the sun $20^\circ$ behind the Earth.  The
triangle is oriented at $60^\circ$ to the ecliptic, so the orbits of
the individual spacecraft will all be in different planes.  This
causes the triangle to spin around itself as it orbits the sun.
Following Cutler, we define a barred ``barycenter'' coordinate system
$(\bar{x},\bar{y},\bar{z})$, which is fixed in space with the
$\bar{x}\bar{y}$-plane aligned with the ecliptic, and an unbarred
``detector'' coordinate system $(x,y,z)$, which is attached to the
detector.  The $z$ axis always points toward the sun, $60^\circ$ away
from vertical, while the $x$ and $y$ axes pinwheel around it.  A
particular binary will have fixed coordinates in the barycenter
system, but its detector coordinates will be time varying.

The three arms act as a pair of two-arm detectors.  We are first
interested in the strain measured in detector I, that formed by arms 1
and 2:
\begin{equation}
h_\mathrm{I}(t) = \frac{\delta L_1(t) - \delta L_2(t)}{L} \, ,
\label{eq:hIpart1}
\end{equation}
where $\delta L_1(t)$ and $\delta L_2(t)$ are the differences in
length in arms 1 and 2 as the wave passes.  $L$ is the unperturbed
length of the arms.  Using the geometry of the detector and the
equation of geodesic deviation \cite{fh05}, we find
\begin{equation}
h_\mathrm{I}(t) = \frac{\sqrt{3}}{2}\left[\frac{1}{2}(h_{xx}-h_{yy})\right] \, .
\label{eq:hIpart2}
\end{equation}
To obtain $h_{xx}$ and $h_{yy}$ for use in these equations, we must
rotate the waveform from the principal axes into the detector frame.
The result is that detector I measures both polarizations, modulated
by the antenna pattern of that detector:

\begin{widetext}
\begin{equation}
h_\mathrm{I}(t) = \frac{\sqrt{3}}{2}\frac{{\mathcal M}^{5/3}(\pi
f)^{2/3}}{D_L}(2[1+(\mathbf{\hat{L}}\cdot
\mathbf{\hat{n}})^2]F^+_\mathrm{I}(\theta_N,\phi_N,\psi_N)\cos[\Phi(t)+\delta_p
\Phi(t)] - 4(\mathbf{\hat{L}}\cdot
\mathbf{\hat{n}})F^{\times}_\mathrm{I}(\theta_N,\phi_N,\psi_N)\sin[\Phi(t)+\delta_p
\Phi(t)]) \, .
\label{eq:hIquadrature}
\end{equation}
\end{widetext}
Detector I acts like a ``standard'' $90^\circ$ GW interferometer
(e.g.\ LIGO), with the response scaled by $\sqrt{3}/2$ (due to the
$60^\circ$ opening angle of the constellation).  The antenna pattern
functions are given by
\begin{align}
\begin{split}
F^+_\mathrm{I}(\theta_N,\phi_N,\psi_N) &= \frac{1}{2}(1+\cos^2\theta_N)\cos
2\phi_N \cos 2\psi_N \\
& \quad - \cos \theta_N \sin 2\phi_N \sin 2\psi_N \, , 
\end{split}
\label{eq:FI+}\\
\begin{split}
F^{\times}_\mathrm{I}(\theta_N,\phi_N,\psi_N) &=
\frac{1}{2}(1+\cos^2\theta_N)\cos 2\phi_N \sin 2\psi_N \\
& \quad +\cos \theta_N \sin 2\phi_N \cos 2\psi_N \, ,
\end{split}
\label{eq:FIx} 
\end{align}
where $\theta_N$ and $\phi_N$ are the spherical angles for the
binary's direction $\mathbf{\hat{n}}$ in the (unbarred) detector frame
and $\psi_N$ is the polarization angle of the wave in that frame:
\begin{equation}
\tan \psi_N = \frac{\mathbf{\hat{q}}\cdot
\mathbf{\hat{z}}}{\mathbf{\hat{p}}\cdot \mathbf{\hat{z}}} =
\frac{\mathbf{\hat{L}}\cdot \mathbf{\hat{z}}-(\mathbf{\hat{L}}\cdot
\mathbf{\hat{n}})(\mathbf{\hat{z}}\cdot
\mathbf{\hat{n}})}{\mathbf{\hat{n}}\cdot (\mathbf{\hat{L}}\times
\mathbf{\hat{z}})} \, .
\label{eq:psiS}
\end{equation}
In order to use these expressions, we must relate the time-dependent
angles in the unbarred detector frame, $\theta_N, \phi_N,$ and
$\psi_N$, to quantities in the barred barycenter frame.  Again, the
details can be found in Cutler \cite{c98}.

Similar expressions hold for detector II.  Following Cutler, we
construct the signal from detector II as
\begin{equation}
h_{\mathrm{II}}(t) = \frac{1}{\sqrt{3}}\left[h_\mathrm{I}(t) + 2h_{\mathrm{II}^\prime}(t)\right] \, , 
\label{eq:detectorII}
\end{equation}
where $h_\mathrm{I}$ is the signal from detector I \eqref{eq:hIpart1},
and $h_{\mathrm{II}^\prime} = (\delta L_2(t)-\delta L_3(t))/L$ is the
signal formed from the difference in the lengths of arms 2 and 3.
This choice makes the noise in detector I uncorrelated with the noise
in detector II; we will exploit this property in Sec.\
{\ref{sec:meas}} to treat detectors I and II as independent detectors.
From \eqref{eq:detectorII}, we obtain
\begin{equation}
h_{\mathrm{II}}(t) = \frac{\sqrt{3}}{2}\left[\frac{1}{2}(h_{xy}+h_{yx})\right]
\, .
\label{eq:hII}
\end{equation}
The result is that detector II also behaves like a $90^\circ$
interferometer (scaled by $\sqrt{3}/2$), but rotated by $45^\circ$
with respect to detector I.  Thus the antenna patterns for detector II
are
\begin{align}
F^+_{\mathrm{II}}(\theta_N,\phi_N,\psi_N) &=
F^+_\mathrm{I}(\theta_N,\phi_N-\pi/4,\psi_N) \, ,
\label{eq:FII+}\\
F^{\times}_{\mathrm{II}}(\theta_N,\phi_N,\psi_N) &=
F^{\times}_\mathrm{I}(\theta_N,\phi_N-\pi/4,\psi_N) \, .
\label{eq:FIIx}
\end{align}
We now rewrite the waveform in terms of an amplitude and phase.
Letting $i = \mathrm{I}, \mathrm{II}$ label detector number, the
waveform as measured by detector $i$ is
\begin{equation}
\begin{split}
h_i(t) &= 2 \frac{{\mathcal M}^{5/3}(\pi
f)^{2/3}}{D_L}A_{\mathrm{pol}, i}(t)\cos[\Phi(t) \\
& \quad + \varphi_{\mathrm{pol},i}(t) + \varphi_D(t) + \delta_p
\Phi(t)] \, ,
\end{split}
\label{eq:fullmeassignal}
\end{equation}
where
\begin{equation}
A_{\mathrm{pol}, i}(t) = \frac{\sqrt{3}}{2}[(1+(\mathbf{\hat{L}} \cdot
\mathbf{\hat{n}})^2)^2F_i^+(t)^2 + 4(\mathbf{\hat{L}} \cdot
\mathbf{\hat{n}})^2F_i^{\times}(t)^2]^{1/2} 
\label{eq:polamp}
\end{equation}
is the ``polarization amplitude'' and
\begin{equation}
\varphi_{\mathrm{pol},i}(t) =
\tan^{-1}\left[\frac{2(\mathbf{\hat{L}}\cdot
\mathbf{\hat{n}})F_i^{\times}(t)}{[1+(\mathbf{\hat{L}}\cdot
\mathbf{\hat{n}})^2]F_i^+(t)}\right] \,
\label{eq:polphase}
\end{equation}
is the ``polarization phase'' {\cite{c98}}.  We have introduced the
``Doppler phase'' $\varphi_D(t)$, which arises from the detector's
motion around the sun and is given by
\begin{equation}
\varphi_D(t) = 2\pi f(t) R_{\oplus}
\sin{\bar{\theta}_N}\cos[\bar{\Phi}_D(t)-\bar{\phi}_N] \, ,
\label{eq:dopphase}
\end{equation}
where $\bar{\Phi}_D(t)$ is the orbital phase of the detector and
$R_{\oplus} = 1$ AU.

Much of our analysis is done in the frequency domain.  We define the
Fourier transform of the signal as
\begin{equation}
\tilde{h}(f) = \int_{-\infty}^{\infty} e^{2\pi i ft}h(t) dt \, .
\end{equation}
To evaluate the Fourier transform, we make use of the stationary phase
approximation \cite{cf94,pw95}.  This approximation relies on the fact
that the orbital time scale $T_{\mathrm{orb}}$ is much shorter than
the precession time scale $T_{\mathrm{prec}}$, as well as the inspiral
time scale $T_{\mathrm{insp}}$ and detector orbital time scale $T_D =
1$ yr.  The result thus differs from the true Fourier transform by
terms of order $T_{\mathrm{orb}}/T_{\mathrm{prec}}$ and
$T_{\mathrm{orb}}/T_{\mathrm{insp}}$ \cite{acst94}.  The Fourier
transform is thus likely to be inaccurate near the end of the
inspiral, when all of these time scales become comparable.  Using
\eqref{eq:PNtime} and \eqref{eq:PNphase}, we have
\begin{equation}
\begin{split}
\tilde{h}_i(f) &= \sqrt{\frac{5}{96}}\frac{\pi^{-2/3} {\mathcal
M}^{5/6}}{D_L}A_{\mathrm{pol}, i}[t(f)]f^{-7/6} \\
& \quad \times e^{i(\Psi(f) - \varphi_{\mathrm{pol},i}[t(f)] -
\varphi_D[t(f)] - \delta_p\Phi[t(f)])} \, ,
\label{eq:freqdomainsignal}
\end{split}
\end{equation}
where the phase $\Psi(f)$ is given by
\begin{equation}
\begin{split}
\Psi(f) &= 2\pi f t_c - \Phi_c - \frac{\pi}{4} + \frac{3}{128}(\pi
{\mathcal M} f)^{-5/3}\\
& \quad \times\left[1 + \frac{20}{9}\left(\frac{743}{336} +
\frac{11}{4}\eta\right)(\pi Mf)^{2/3}\right.\\
& \quad - 4(4\pi - \beta)(\pi Mf) + 10\left(\frac{3058673}{1016064} +
\frac{5429}{1008}\eta \right.\\
& \quad \left.\left. + \frac{617}{144}\eta^2 - \sigma \right)(\pi
Mf)^{4/3}\right] \, .
\end{split}
\label{eq:PNpsi}
\end{equation}
In the work by Cutler \cite{c98}, the separation of time scales that
we used above leads to an interpretation of the polarization
amplitude, polarization phase, and Doppler phase as modulations, in
amplitude and phase, of an underlying carrier signal.  These
modulations make it possible to measure the sky position of the
source, which also helps to measure the luminosity distance $D_L$
\cite{h02}.  With the addition of precession, the polarization
amplitude and polarization phase include additional modulations which
further improve the measurement of these parameters.  In conjunction
with the purely intrinsic effects of precession, these effects also
help us to better measure the masses and spins of the system.

\section{Measurement and parameter estimation with LISA}
\label{sec:meas}

\subsection{Theory}
\label{sec:meastheory}

In the previous section, we constructed the expected form for the GW
strain that LISA is being designed to measure.  The signal $s_i(t)$ as
measured by detector $i$ will of course also include noise $n_i(t)$:
\begin{equation}
s_i(t) = h_i(t) + n_i(t) \, .
\label{eq:signalplusnoise}
\end{equation}
The LISA noise spectrum is discussed in section \ref{sec:lisanoise};
in this section, we discuss the theory of parameter estimation with a
noisy signal.  First, consider only one detector.  We assume that the
noise is zero mean, wide-sense stationary, and Gaussian.  Wide-sense
stationary means that the autocovariance function
\begin{equation}
K_{n}(t,t') = \langle n(t)n(t')\rangle-\langle n(t)\rangle\langle
n(t')\rangle
\label{eq:autocovariance}
\end{equation}
depends only on the time difference $\tau = t-t'$.  (Throughout this
section, quantities within angle brackets are ensemble averaged with
respect to the noise distribution.)  A process is Gaussian if every
sample of the process can be described as a Gaussian random variable
and all possible sets of samples of the process are jointly Gaussian.
However, the noise is colored, not white.  A white noise process is
defined to be a process which is uncorrelated with itself at different
times; that is, its autocovariance is a delta function.  Because the
noise is colored, it has an interesting (nonflat) power spectral
density (PSD), which is defined as the Fourier transform of the
autocovariance function:

\begin{equation}
S_{n}(f) = 2\int_{-\infty}^{\infty}d\tau e^{2\pi if\tau}K_{n}(\tau) \, .
\label{eq:PSD}
\end{equation}
The factor of 2 follows \cite{cf94}; we actually use the {\em
one-sided} PSD.  Since the noise is Gaussian, it is described entirely
by its second moments.  Therefore, we will only need the PSD, and not
the full probability density function, to analyze the effect of the
noise on the signal.

Incidentally, it can be shown that wide sense stationarity implies
that the Fourier transform of $n(t)$ is a nonstationary white noise
process in frequency:
\begin{equation}
\langle\tilde{n}(f)\tilde{n}^*(f')\rangle = \frac{1}{2}\delta(f-f')S_n(f) \, .
\label{eq:fouriercorrelation}
\end{equation}
The Fourier components are thus independent Gaussian random variables.

Now briefly consider both detectors.  We explicitly constructed the
second detector \eqref{eq:detectorII} [with $h(t) \rightarrow s(t)$]
so that the noise in it is uncorrelated with, and thus independent of,
noise in the first detector.  Thus we have
\begin{equation}
\langle\tilde{n}_i(f)\tilde{n}_j^*(f)\rangle =
\frac{1}{2}\delta_{ij}\delta(f-f')S_n(f) \, .
\label{eq:fouriercorrelation2}
\end{equation}
The uncorrelated nature of these two noises will allow us to easily
generalize discussion from one detector to the full two effective
detector system.

Let us write our generalized GW as $h(\boldsymbol{\theta})$, where the
components of the vector $\boldsymbol{\theta}$ represent the various
parameters on which the waveform depends.  We now assume that a GW
signal with particular parameters $\boldsymbol{\tilde\theta}$ is
present in the data (i.e., ``detection'' has already occurred), and
want to obtain estimates $\boldsymbol{\hat{\theta}}$ of those source
parameters.  Finn \cite{f92} shows that the probability for the noise
to have some realization $n_0(t)$ is given by
\begin{equation}
p(n=n_0) \propto e^{-(n_0|n_0)/2} \,,
\label{eq:noiseprob}
\end{equation}
where the inner product used here is given by
\begin{align}
(a|b) &= 4 \, \mathrm{Re} \int_0^{\infty} df
\frac{\tilde{a}^*(f)\tilde{b}(f)}{S_n(f)} \\
&= 2 \int_0^{\infty} df \frac{\tilde{a}^*(f)\tilde{b}(f) +
\tilde{a}(f)\tilde{b}^*(f)}{S_n(f)} \, .
\label{eq:innerproduct}
\end{align}
This product is a natural one for the vector space of (frequency
domain) signals $a(f)$.  (Note that this definition of the inner
product differs from \cite{f92} by a factor of 2.)

Given a particular measured signal $s(t)$, the probability that the GW
parameters are given by $\boldsymbol{\tilde{\theta}}$ is the same as
the probability that the noise takes the realization $s -
h(\boldsymbol{\tilde{\theta}})$:
\begin{equation}
p(\boldsymbol{\tilde{\theta}}|s) \propto
e^{-(h(\boldsymbol{\tilde{\theta}}) - s |
h(\boldsymbol{\tilde{\theta}}) - s)/2} \, ,
\label{eq:signalprob}
\end{equation}
where the constant of proportionality may include prior probability
densities for the parameters $\boldsymbol{\tilde{\theta}}$.  For
simplicity, we take these to be uniform.

We can estimate the parameters $\boldsymbol{\tilde{\theta}}$ by the
maximum likelihood (ML) method.  This method involves finding the
parameters $\boldsymbol{\hat{\theta}}$ that maximize
(\ref{eq:signalprob}), or alternatively, minimize
$(h(\boldsymbol{\tilde{\theta}}) - s | h(\boldsymbol{\tilde{\theta}})
- s)$, which can be considered a distance in signal space.  A bank of
template waveforms is correlated with the received signal and,
assuming that any template produces a statistically significant
correlation, the one with the highest correlation is the one with the
ML parameters.  The SNR for this signal is then given by \cite{cf94}

\begin{equation}
\rho \approx (h(\boldsymbol{\hat{\theta}}) |
h(\boldsymbol{\hat{\theta}}))^{1/2} \approx
(h(\boldsymbol{\tilde{\theta}}) |
h(\boldsymbol{\tilde{\theta}}))^{1/2} \, .
\label{eq:finalSNR}
\end{equation}
To quantify the errors in the ML estimate, we expand
\eqref{eq:signalprob} around the most likely values
$\boldsymbol{\hat{\theta}}$.  We can then write the probability
density as \cite{cf94, pw95}:

\begin{equation}
p(\boldsymbol{\tilde{\theta}} | s) \propto e^{-\Gamma_{ab}\delta
\theta^a \delta \theta^b/2} \, ,
\label{eq:linearizedprob}
\end{equation}
where $\delta \theta^a = \tilde{\theta}^a - \hat{\theta}^a$ and

\begin{equation}
\Gamma_{ab} = \left(\frac{\partial h}{\partial \theta^a}\left|
\frac{\partial h}{\partial \theta^b}\right.\right) \, ,
\label{eq:fishermatrix}
\end{equation}
evaluated at $\boldsymbol{\theta} = \boldsymbol{\hat{\theta}}$, is the
Fisher information matrix.  For small deviations from the ML estimate,
the distribution is Gaussian.  This expression holds for large values
of the SNR \eqref{eq:finalSNR}.  It is worth emphasizing at this point
that, in our evaluation of Eq.\ (\ref{eq:fishermatrix}), most
derivatives are taken numerically using finite differencing --- the
complicated nature of the signal (due to the inclusion of spin
precession) makes it essentially impossible to evaluate all but a few
of our derivatives analytically.  This is another reason that the code
we have developed for this analysis is substantially slower than those
developed for analyses which do not include spin-precession physics.

Now we return again to the two detector case.  Using
\eqref{eq:fouriercorrelation2}, we can write a total Fisher matrix as
the sum of the individual Fisher matrices for each detector:
\begin{equation}
\Gamma_{ab}^{\rm tot} = \Gamma_{ab}^\mathrm{I}+\Gamma_{ab}^{\mathrm{II}} \, .
\label{eq:fishertot}
\end{equation}
The Fisher matrix is then inverted to produce the covariance matrix
$\Sigma^{ab} = (\Gamma_{\rm tot}^{-1})^{ab}$.  The diagonal terms of
the covariance matrix represent measurement errors:

\begin{equation}
\Delta \theta^a \equiv \sqrt{\langle(\delta \theta^a)^2\rangle} =
\sqrt{\Sigma^{aa}} \, .
\label{eq:error}
\end{equation}
The off-diagonal terms can be expressed as correlation coefficients,
ranging from $-1$ to $1$:
\begin{equation}
c^{ab} \equiv \frac{\langle\delta \theta^a \delta
\theta^b\rangle}{\Delta \theta^a \Delta \theta^b} =
\frac{\Sigma^{ab}}{\sqrt{\Sigma^{aa}\Sigma^{bb}}} \, .
\label{eq:correlation}
\end{equation}

\subsection{LISA detector and astrophysical noise}
\label{sec:lisanoise}

We turn now to a discussion of the noise we expect in LISA
measurements.  Our model for the instrumental noise spectrum,
$S_h^{\rm inst}(f)$, is based on that described in Ref.\ \cite{lhh00}.
(From now on, we use the notation $S_h$ for strain noise instead of
$S_n$ for general noise.)  In particular, we use the online
sensitivity curve generator provided by S.\ Larson, which implements
the recipe of {\cite{lhh00}} (see \cite{sensitivity}).  The output of
Larson's webtool gives a sky averaged {\it amplitude} sensitivity
curve, $h_\mathrm{Larson}$.  To convert to the noise we need for our
analysis, we square this amplitude and insert two numerical factors:
\begin{equation}
S_h^{\rm inst}(f) =
\frac{1}{5}\times\left(\frac{\sqrt{3}}{2}h_\mathrm{Larson}\right)^2 =
\frac{3}{20}h_\mathrm{Larson}^2\;.
\label{eq:convert_inst_noise}
\end{equation}
The factor of $1/5$ accounts for the averaging of the antenna pattern
functions over all sky positions and source orientations.  This factor
is only correct for measuring radiation with wavelength $\lambda \gg
L$ (where $L$ is the LISA arm length).  As a consequence, our
instrumental noise will be inaccurate at high frequencies.  This will
have little impact on our analysis since, as already argued, the
signal from merging binary black holes accumulates at low frequencies.

The factor $\sqrt{3}/2$ arises due to the $60^\circ$ opening angle of
the interferometer arms; we have already accounted for this factor in
our discussion of the interferometer's interaction with a GW [cf.\
Eqs.\ (\ref{eq:hIpart2}) and (\ref{eq:hII})].  The numerical factor
$3/20$ has been the source of some confusion; Berti, Buonanno and Will
very nicely straightened this out.  See Sec.\ IIC of Ref.\
{\cite{bbw05}} for further discussion of these factors.

Besides purely instrumental noise, LISA data will contain ``noise''
from a background of confused binary sources\footnote{While surely
noise when studying cosmological black holes, this background is {\it
signal} to those interested in stellar populations.}, mostly white
dwarf binaries.  An isotropic background of indistinguishable sources
can be represented as noise with spectral density \cite{bc04}

\begin{equation}
S^{\mathrm{conf}}_h(f) = \frac{3}{5\pi} f^{-3}\rho_c \Omega_{GW}(f) \, , 
\end{equation}
where $\rho_c = 3H_0^2/8\pi$ is the critical energy density to close
the universe and $\Omega_{GW} = (f/\rho_c)d\rho_{GW}/df$ is the energy
density in GWs relative to $\rho_c$ per logarithmic frequency
interval.  Using this form and the results of Farmer and Phinney
\cite{fp03}, we model the confusion noise due to extragalactic binary
sources by

\begin{equation}
S^{\mathrm{exgal}}_h(f) = 4.2 \times 10^{-47} \left(\frac{f}{1 \,
\mathrm{Hz}}\right)^{-7/3} \mathrm{Hz}^{-1} \, .
\label{eq:Sexgal}
\end{equation}
From Nelemans {\it et al.}\ \cite{nyz01}, we take the galactic white
dwarf confusion noise to be
\begin{equation}
S^{\mathrm{gal}}_h(f) = 2.1 \times 10^{-45} \left(\frac{f}{1 \,
\mathrm{Hz}}\right)^{-7/3} \mathrm{Hz}^{-1} \, .
\label{eq:Sgal}
\end{equation}
The combined instrumental and galactic confusion noise is given by
\cite{bc04}
\begin{equation}
\begin{split}
S^{\mathrm{inst+gal}}_h(f) &= \min
[S^{\mathrm{inst}}_h(f)/\exp(-\kappa T_{\mathrm{mission}}^{-1}dN/df),
\\
& \quad S^{\mathrm{inst}}_h(f) + S^{\mathrm{gal}}_h(f)] \, .
\end{split}
\label{eq:Sinst+gal}
\end{equation}
The choice taken in Eq.\ (\ref{eq:Sinst+gal}) reflects the fact that,
at sufficiently high frequency, the number of binaries per bin should
be small enough that they are no longer truly confused and can be
subtracted from the data stream (at least partially).  The factor
$\exp (-\kappa T_{\mathrm{mission}}^{-1} dN/df)$ is the fraction of
``uncorrupted'' frequency bins.  We choose $\kappa = 4.5$ \cite{c03},
$T_{\mathrm{mission}}$ is the mission duration (which we take to be
three years), and
\begin{equation}
\frac{dN}{df} = 2 \times 10^{-3} \left(\frac{1 \,
\mathrm{Hz}}{f}\right)^{11/3} {\mathrm{Hz}}^{-1} \,
\label{eq:dNdf}
\end{equation}
is the number density of galactic binaries per unit frequency
\cite{h02}.

Finally, the total noise is given by
\begin{equation}
S_h(f) = S^{\mathrm{inst + gal}}_h(f) + S^{\mathrm{exgal}}_h(f) \, .
\label{eq:Sh}
\end{equation}

\section{Results}
\label{sec:results}

\subsection{Procedure}
\label{sec:procedure}

\subsubsection{Parameter space}

Seventeen parameters describe the most general binary black hole
inspiral waveform \cite{v04}.  Two of these are the orbital
eccentricity and the orientation of the orbital ellipse; since we only
consider circular orbits, we can ignore these two.  The other 15
parameters are all necessary to describe the full post-Newtonian
waveform with precession effects that we described in section
\ref{sec:inspiral}.

We divide this set into intrinsic and extrinsic parameters.  In our
labeling system, intrinsic parameters are those which label properties
intrinsic to the binary itself; extrinsic parameters label properties
which depend upon the position and placement of the binary relative to
the observer.  One can regard intrinsic parameters as describing the
physics or astrophysics of the binary system, and extrinsic parameters
as describing the binary's astronomical properties.

The intrinsic parameters we use are $\ln m_1$; $\ln m_2$;
$\bar{\mu}_{L}(0) \equiv \cos[\bar{\theta}_{L}(0)]$ and
$\bar{\phi}_{L}(0)$, the initial direction of the orbital angular
momentum; $\bar{\mu}_{S_1}(0) \equiv \cos[\bar{\theta}_{S_1}(0)]$,
$\bar{\phi}_{S_1}(0)$, $\bar{\mu}_{S_2}(0) \equiv \cos
[\bar{\theta}_{S_2}(0)]$, and $\bar{\phi}_{S_2}(0)$, the initial
directions of the spins; $\chi_1$ and $\chi_2$, the dimensionless spin
parameters; $t_c$, the time at coalescence; and $\Phi_c$, the phase at
coalescence.  (Note that $t_c$ and $\Phi_c$ could very well be
considered extrinsic in our labeling system, since they just label the
system's state at some particular time.  At any rate, neither $t_c$
nor $\Phi_c$ is of much physical interest, so their categorization
isn't too important.)  Our extrinsic parameters are $\bar{\mu}_N =
\cos \bar{\theta}_N$ and $\bar{\phi}_N$, the sky position in
barycenter coordinates; and $\ln D_L$, the luminosity distance to the
binary.  All of these parameters must be fit in a measurement and thus
must be included in our Fisher matrix analysis.  We are not
necessarily interested in all of them, however.  In particular, we
will focus on the masses, the dimensionless spin parameters, the sky
position, and the luminosity distance.

It is worth noting that this choice of parameters is not the same as
that used in analyses which neglect precession.  In that case, the
direction of the angular momentum $\mathbf{\hat{L}}$ is constant, and
so the system's orientation is constant and fully described using two
angles (e.g., $\bar{\mu}_L$ and $\bar{\phi}_L$).  Including
precession, $\mathbf{\hat{L}}$ is no longer constant, but evolves
according to \eqref{eq:Ldot}.  The solution to this differential
equation requires two initial conditions, for instance,
$\bar{\mu}_L(0)$ and $\bar{\phi}_L(0)$, which can be used as
parameters of the system.  Since these initial conditions are taken at
the (somewhat arbitrary) starting point of our calculations, they do
not hold much physical interest (though they must be fit for and thus
included in our Fisher matrix).

Previous analyses, including the precursor to this work \cite{h02},
have used $\beta$ \eqref{eq:beta} and $\sigma$ \eqref{eq:sigma} as
parameters --- these are constants when precession is neglected.  They
are also the only combinations of the spin magnitudes and spin angles
that enter into the expression for the waveform.  Boiling the six
numbers which characterize $\mathbf{S}_1$ and $\mathbf{S}_2$ down to
two greatly simplifies the parameter space, but also restricts us from
being able to measure, for example, the black holes' spin magnitudes.
When precession is included, $\beta$ and $\sigma$ are no longer
constants.  In addition, they no longer fully characterize the signal,
since the precession equations \eqref{eq:S1dot}, \eqref{eq:S2dot}, and
\eqref{eq:Ldot} depend on all of the components of the spins. We thus
need six spin-related parameters to fully describe the signal: the
magnitudes of the spins and their orientations at some initial time.
The orientations are again uninteresting, but the fact that we can
measure the magnitudes of the spins and quantify their errors is quite
interesting and new to this analysis.

Finally, we break from tradition and use $\ln m_1$ and $\ln m_2$ to
parameterize our masses rather than $\ln {\mathcal M}$ and $\ln \mu$.
The chirp mass and reduced mass have been used in most previous work
because of their appearance in the waveform phase $\Psi(f)$.  However,
the precession equations, as well as the spin parameters $\beta$ and
$\sigma$, depend on the individual masses of the black holes.  It is a
simple matter in principle to just solve for $m_{1,2}({\cal M},\mu)$
and substitute into the precession equations.  Unfortunately, the
Jacobian of the transformation between $({\cal M},\mu)$ and
$(m_1,m_2)$ is singular when $m_1 = m_2$, leading to problems in
evaluating the Fisher matrix.

These problems can be illustrated analytically.  Consider how
derivatives of some function $f(m_1,m_2)$ with respect to ${\cal M}$
behave:

\begin{equation}
\frac{\partial f}{\partial {\mathcal M}} = \sum_{i=1}^2 \frac{\partial
f(m_1,m_2)}{\partial m_i}\frac{\partial m_i(\mathcal{M},\mu)}{\partial
{\mathcal M}} \, .
\label{eq:infinitederiv}
\end{equation}
When $m_1 = m_2$, the final derivative diverges --- a behavior that we
have seen numerically.  The Fisher information is infinite, and the
Gaussian approximation breaks down; the same problem occurs for $\mu$.
Thus, we argue that, when precession is included, $\mathcal{M}$ and
$\mu$ are no longer a good choice of parameters to describe the
system.  Since we are still interested in the errors in $\ln {\mathcal
M}$ and $\ln \mu$ (which are determined to higher accuracy than the
individual masses), we convert using the propagation of errors
formulas

\begin{eqnarray}
\left(\frac{\Delta {\mathcal M}}{{\mathcal M}}\right)^2 &=&
\left(\frac{m_1}{{\mathcal M}}\right)^2\left(\frac{\partial{\mathcal
M}}{\partial m_1}\right)^2\left(\frac{\Delta m_1}{m_1}\right)^2
\nonumber\ \\
&+& \left(\frac{m_2}{{\mathcal
M}}\right)^2\left(\frac{\partial{\mathcal M}}{\partial
m_2}\right)^2\left(\frac{\Delta m_2}{m_2}\right)^2
\nonumber\\
&+& 2\left(\frac{m_1m_2}{{\mathcal
M}^2}\right)\left(\frac{\partial{\mathcal M}}{\partial
m_1}\right)\left(\frac{\partial{\mathcal M}}{\partial
m_2}\right)\Sigma^{\ln m_1, \ln m_2} \, ,
\nonumber\\
\label{eq:m1m2toMc}
\end{eqnarray}
\begin{eqnarray}
\left(\frac{\Delta \mu}{\mu}\right)^2 &=&
\left(\frac{m_1}{\mu}\right)^2\left(\frac{\partial \mu}{\partial
m_1}\right)^2\left(\frac{\Delta m_1}{m_1}\right)^2
\nonumber\\
&+& \left(\frac{m_2}{\mu}\right)^2\left(\frac{\partial
\mu}{\partial m_2}\right)^2\left(\frac{\Delta m_2}{m_2}\right)^2
\nonumber\\
&+& 2\left(\frac{m_1m_2}{\mu^2}\right)\left(\frac{\partial
\mu}{\partial m_1}\right)\left(\frac{\partial\mu}{\partial
m_2}\right)\Sigma^{\ln m_1, \ln m_2} \, .
\nonumber\\
\label{eq:m1m2tomu}
\end{eqnarray}
For unequal masses, we find that computing errors in $m_1$ and $m_2$
and then converting gives the same result as simply computing errors
in ${\cal M}$ and $\mu$ directly.  We do not find good agreement in
the equal mass case; for the reasons discussed above, however, we do
not trust the (${\cal M},\mu$) parameterization in this case.  At any
rate, the case $m_1 = m_2$ is quite implausible in nature, so this is
almost certainly a moot point as far as real measurements are
concerned\footnote{It is worth noting that even a slight mass
difference (a few percent) is sufficient for the two approaches to
match.}.  We note that Vecchio \cite{v04}, for simplicity, considers
the equal mass case exclusively but does not report any anomalous
behavior such as we have seen.  We are puzzled about this discrepancy.

\subsubsection{Calculations}

The code we use to calculate parameter measurement errors is based on
that used in \cite{h02}.  It is written in C++ using several routines
taken, sometimes with slight modification, from \cite{numrec}.  As in
\cite{h02}, we perform Monte Carlo simulations in which we specify
rest-frame masses and redshift and then randomly choose sky position,
initial angular momentum and spin directions, spin magnitudes, and
time of coalescence within the three-year LISA mission window.  We
specify spin magnitudes for some studies as well.

The primary function of the code is the calculation of the full
gravitational waveform, including precessional effects.  In order to
effectively use the formulas of section \ref{sec:meas}, we take the
wave frequency $f$ as the independent variable.  The elapsed time is
related to the frequency using \eqref{eq:PNtime}.  The calculation is
started when the waveform enters LISA's band (taken to be
$f_{\mathrm{min}} = 3\times 10^{-5}$ Hz throughout this paper) or when
the LISA mission begins, whichever is later.  (By treating the time of
coalescence as a Monte Carlo variable, some signals will be partially
cut off because they are already in band when LISA begins
observations.)  The calculation proceeds until the binary reaches the
Schwarzschild innermost stable circular orbit (ISCO) at orbital
separation $r = 6M$.  Though perhaps a somewhat crude choice, we use
this criterion for simplicity.  Choosing slightly different cutoff
radii does not change our results very much; at any rate, the
post-Newtonian phase formula and precession equations we use in this
regime are unlikely to be very accurate.  The frequency at this point,
which we call the ``merge frequency,'' can be found using
\eqref{eq:PNomega} for $r = 6M$ (plus $f = \Omega/\pi$).  For
simplicity, we just use the lowest order term: $f =
(M/r^3)^{1/2}/\pi$.  Any error due to this approximation is no doubt
unimportant compared to the arbitrary selection of $r = 6M$ for the
transition.

Once the frequency range has been determined, the true work begins.
We integrate the precession equations \eqref{eq:S1dot},
\eqref{eq:S2dot}, and \eqref{eq:Ldot} using a Runge-Kutta routine to
find the values of $\mathbf{\hat{L}}$, $\mathbf{\hat{S}}_1,$ and
$\mathbf{\hat{S}}_2$ over the duration of the signal.  The routine is
a fifth-order adaptive step algorithm in the frequency domain.  At
each frequency, the code takes the results for the three orbital
angular momentum components and six spin components and uses them to
calculate $\bar{\mu}_L$, $\bar{\phi}_L$, $\beta$, and $\sigma$.  It
also computes the integrated correction to the phase using the
derivative (\ref{eq:dTomphase}).

As already discussed, our derivatives are taken numerically rather
than analytically.  We therefore must do the integration described
above a total of 21 times: once for the given values of the
parameters, and twice more for small shifts in each parameter which
requires a numerical derivative.  This repetition slows the code quite
drastically compared to its earlier incarnation --- an unfortunate but
unavoidable cost.

Once all of the necessary integrations are complete, the SNR
\eqref{eq:finalSNR} and the Fisher matrix \eqref{eq:fishermatrix} can
be calculated for each of the two effective detectors of LISA using
the noise $S_h(f)$ (see Sec.\ {\ref{sec:lisanoise}}).  Some previous
work \cite{c98,bbw05} investigated parameter estimation using the
signal from only one synthesized detector; we will always assume that
both are operational.  It would be interesting to see how measurement
degradation due to only having a single operating detector can be
ameliorated by including precession effects.

At this stage, the necessary integrals are performed using
Curtis-Clenshaw quadrature, which depends on the decomposition of the
integrand into Chebyshev polynomials \cite{numrec}.  This method keeps
the code reasonably fast even with the addition of the Runge-Kutta
routine.  At each step of the integration, the integrator uses the
values that were calculated using that Runge-Kutta routine to evaluate
the waveform and/or its appropriate derivatives.  The derivatives are
calculated using
\begin{equation}
\frac{df}{d\theta} \approx \frac{f(\theta+\frac{\Delta
\theta}{2})-f(\theta-\frac{\Delta \theta}{2})}{\Delta \theta} \, .
\label{eq:numderiv1}
\end{equation}
For all parameters, we use $\Delta \theta = 10^{-5} \, \theta$.  We
invert the Fisher matrix using LU decomposition to produce the
covariance matrix {\cite{numrec}}.  In ``poor'' cases, (e.g., high
mass binaries at large redshift) the Fisher matrix can be nearly
singular, with a large condition number\footnote{The ``condition
number'' is the ratio of the largest eigenvalue of a matrix to the
smallest.  A rule of thumb is that matrix inversion breaks down when
the logarithm of the condition number of a matrix exceeds the number
of digits of accuracy in the matrix elements (see, e.g., discussion in
{\cite{numrec}}).}.  In such a case, the covariance matrix produced by
the code may not be the true inverse of the Fisher matrix (and may not
even be positive definite).  This problem is largely ameliorated by
representing our numerical data in {\tt long double} format --- this
improves (relative to type {\tt double}) matrix inverses in many
``bad'' cases but leaves all other cases essentially unchanged.

It is worth noting that the bad cases are typically ones in which the
binary executes very few orbits over the course of the measurement.
We are confident in our results for all cases in which the number of
measured orbits, $N_{\rm orb} = \Phi_{\rm orb}/2\pi$, is greater than
$\sim 10 - 20$.  When the number of orbits is small (and the condition
number is concomitantly high), the errors are so large that they are
basically meaningless.  In such a case, measurement would not
determine the system's characteristics in any meaningful sense.

\subsection{Black hole masses and spins}
\label{sec:resI}

\begin{figure}[htb]
\includegraphics[scale=0.54]{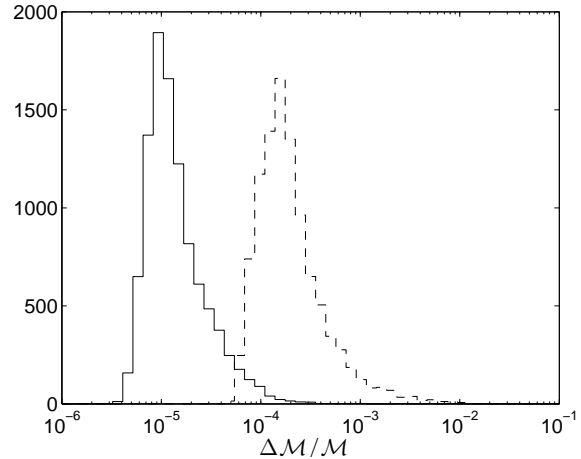}
\caption{Distribution of errors in chirp mass $\mathcal{M}$ for $10^4$
binaries with $m_1 = 10^6 M_\odot$ and $m_2 = 3 \times 10^5 M_\odot$
at $z = 1$.  The dashed line is the precession-free calculation; the
solid line includes precession.  Precession reduces the measurement
error by about an order of magnitude.}
\label{fig:Mccomparison}
\end{figure}

\begin{figure}[htb]
\includegraphics[scale=0.54]{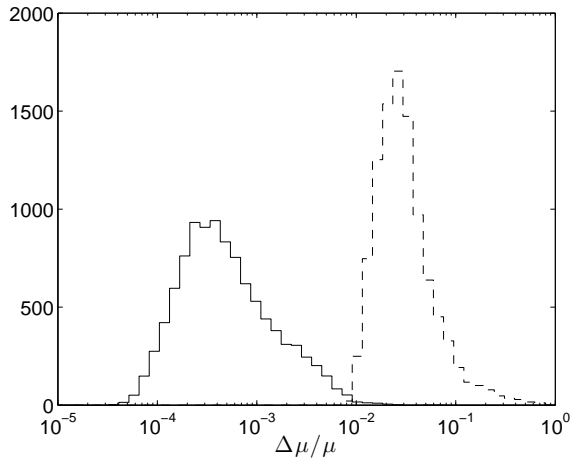}
\caption{Distribution of errors in reduced mass $\mu$ for $10^4$
binaries with $m_1 = 10^6 M_\odot$ and $m_2 = 3 \times 10^5 M_\odot$
at $z = 1$.  The dashed line is the precession-free calculation; the
solid line includes precession.  Precession has an enormous effect on
the reduced mass, which was previously highly correlated with the
parameters $\beta$ and $\sigma$.}
  \label{fig:mucomparison}
\end{figure}

Representative examples of our results are shown in Figures
\ref{fig:Mccomparison} and \ref{fig:mucomparison}.  These histograms
show the spread of errors in $\mathcal{M}$ and $\mu$ for a sample of
$10^4$ binaries at $z = 1$ with rest frame masses $m_1 = 10^6 \,
M_{\odot}$ and $m_2 = 3\times 10^5 \, M_{\odot}$.  Each figure
compares the results of the new code to those of the original code of
\cite{h02}, which neglects precession.  (This code has been updated to
reflect up-to-date models for LISA noise; some minor coding errors
have also been corrected.)  Clearly, including spin precession leads
to a significant improvement in the measurement of these mass
parameters.  The reduced mass $\mu$, in particular, is improved.  This
is because the time variation of $\beta$ and $\sigma$ breaks a near
degeneracy between those terms and $\mu$ in the post-Newtonian phase
\eqref{eq:PNpsi}.  The masses also control the precession rate, as
seen in \eqref{eq:S1dot}, \eqref{eq:S2dot}, and \eqref{eq:Ldot}.
(Recall that, in those equations, $S_i = \chi_i m_i^2$.)  This means
that they now influence the polarization amplitude and polarization
phase; they do not influence those quantities when precession is
neglected.  These precession-induced influences on the waveform make
it possible to determine the masses even more accurately than before.

\begin{figure}[htb]
\includegraphics[scale=0.54]{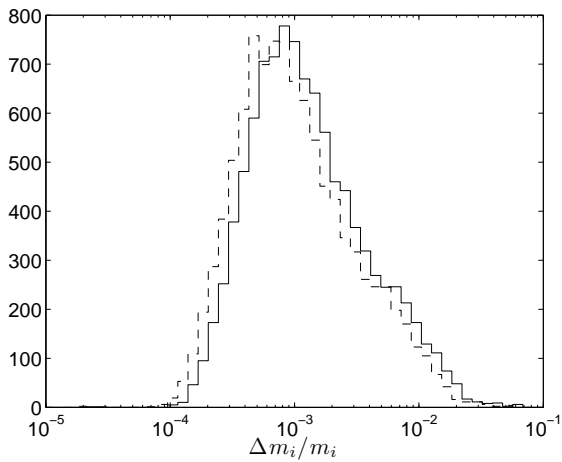}
\caption{Distribution of errors in individual hole masses for $10^4$
binaries at $z = 1$.  The solid line is $m_1 = 10^6 M_\odot$, while
the dashed line is $m_2 = 3 \times 10^5 M_\odot$.  The individual
masses are not determined as well as $\mathcal{M}$ and $\mu$, but they
are better behaved parameters when precession is introduced.}
\label{fig:masses}
\end{figure}

As discussed earlier, we have found the masses $m_1$ and $m_2$ to be
more useful parameters than $\mathcal{M}$ and $\mu$ when precession is
included.  Figure \ref{fig:masses} shows the error in measurements of
the individual masses for our example system.  While these masses are
measured quite accurately, they are not measured as accurately as
$\mathcal{M}$ and $\mu$.  This reflects the fact that, even though the
individual masses play a role in the precession, the other parts of
the waveform depend explicitly on the combinations $\mathcal{M}$ and
$\mu$.  Notice also that the smaller mass is typically determined a
bit better than the larger one, though the difference is not large.

\begin{figure}[htb]
\includegraphics[scale=0.54]{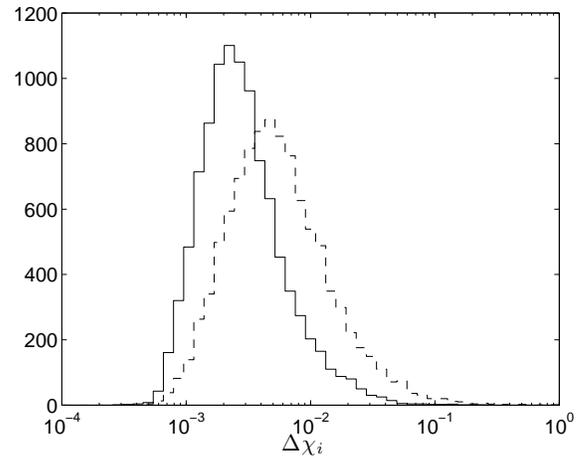}
\caption{Distribution of errors in dimensionless spin parameters
$\chi_1$ (solid line) and $\chi_2$ (dashed line) for $10^4$ binaries
with $m_1 = 10^6 M_\odot$ and $m_2 = 3 \times 10^5 M_\odot$ at $z =
1$.  In each binary, the spin values are randomly selected between 0
and 1.  The higher mass then has, on average, higher total spin and
more effect on the precession.}
\label{fig:spins}
\end{figure}

Precession makes it possible to determine the spins of the binary's
members.  Figure \ref{fig:spins} shows the error in measurements of
the two dimensionless spin parameters $\chi_1$ and $\chi_2$.  We see
that $\chi$ is generally determined very well: Taking a typical spin
parameter to be about 0.5 (recall we randomly choose $\chi$ between 0
and 1), the bulk of this distribution corresponds to errors of a bit
less than a percent.  For this entirely random distribution of $\chi$,
the dimensionless spin parameter of the larger hole tends to be better
determined than that of the smaller hole.  This appears to be a simple
consequence of the fact that black hole spin scales as mass squared
($S_i = \chi_i m_i^2$), and larger spin has more of an impact on the
waveform.

\begin{figure}[htb]
\includegraphics[scale=0.54]{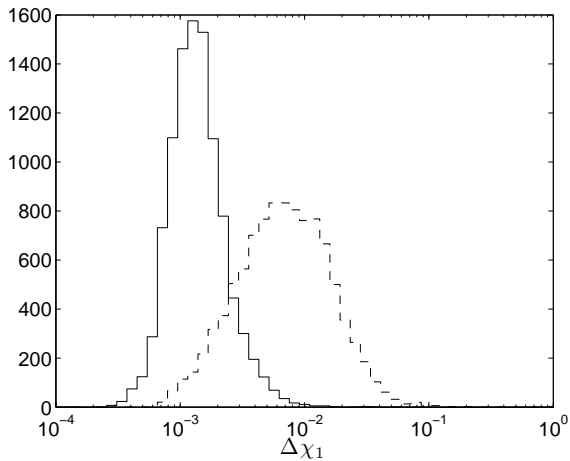}
\caption{Distribution of errors in dimensionless spin parameter
$\chi_1$ for $10^4$ binaries with $m_1 = 10^6 M_\odot$ and $m_2 = 3
\times 10^5 M_\odot$ at $z = 1$.  Here, spin magnitudes have been set
to a specified value --- low spin, $\chi_1 = \chi_2 = 0.1$ (dashed
line), and high spin, $\chi_1 = \chi_2 = 0.9$ (solid line).  Since
greater spin more strongly impacts the waveform, the high spin case is
measured more accurately.}
\label{fig:spinvsspin}
\end{figure}

Next, we examine how well spin is measured as a function of spin
magnitude.  Figure \ref{fig:spinvsspin} shows the error in $\chi_1$
for the same system as in Fig.\ {\ref{fig:spins}}, except that we set
$\chi_1 = \chi_2 = 0.9$ (solid line) and $\chi_1 = \chi_2 = 0.1$
(dashed line), rather than randomly distributing their values.  This
allows us to more accurately assess how well spin is determined as a
function of its value, as well as to more accurately determine the
percent error we expect in these measurements.  For $\chi_1 = \chi_2 =
0.1$, the error is almost $10 \%$, while for $\chi_1 = \chi_2 = 0.9$,
the error is closer to $0.1 \%$.  This is a considerable difference
and is easily ascribed to the fact that rapid spin has a much stronger
impact on the waveform.

Table \ref{table:intrinsic1} shows the median errors in intrinsic
parameters for different mass ratios at $z = 1$.  We continue to
include the errors in $\mathcal{M}$ and $\mu$ for comparison with the
precession-free case, but only in binaries of unequal mass where the
Gaussian approximation is well defined.  Examining the table, we see
some interesting features.  The errors, in general, are worse for
higher mass binaries, which spend less time in the LISA band.  At $m_1
= m_2 = 10^7 \, M_\odot$, the mass errors jump to nearly $10\%$,
compared to tenths of a percent at the next lower mass combination.
In addition, the spin determination becomes very unreliable.  Mass
ratio also has an important effect on the results.  Taking into
account the general trend caused by total mass, we see that unequal
mass ratios generally produce better results.  This is good news for
eventual measurements of astrophysical systems, since merger tree
calculations show that binaries are most likely to have mass ratios of
about 10 \cite{shmv04}.  To understand the mass ratio dependence, we
again turn to the precession equations \eqref{eq:S1dot},
\eqref{eq:S2dot}, and \eqref{eq:Ldot}.  For unequal mass ratios, the
geodetic spin-orbit and spin-spin terms will cause the two spins to
precess at different rates, creating richer features in the signal
than for equal mass ratios.  This illustrates the importance of
effects beyond the ``simple precession'' of \cite{acst94, v04}.  We
also see that the trends of Figs.\ \ref{fig:masses} and
\ref{fig:spins} hold for each unequal mass binary in the table.  That
is, the mass of the smaller hole is determined better than the mass of
the larger hole, but the spin of the larger hole is determined better
than the spin of the smaller hole.

Tables \ref{table:intrinsic3} and \ref{table:intrinsic5} show the same
results for $z = 3$ and $z = 5$, respectively.  The trends we see at
$z = 1$ largely continue at these redshifts.  In general, the errors
get worse at higher redshift as the signal amplitude degrades and more
of the signal is redshifted out of band.  It is worth noting that the
change is generally greater from $z = 1$ to $z = 3$ than from $z = 3$
to $z = 5$.  This effect was also seen by Berti, Buonanno, and Will
\cite{bbw05} and can be explained by considering the redshift
dependence of the wave amplitude.  Neglecting all the angular factors
and remembering to redshift quantities with the dimensions of time, we
find that the amplitude scales like $(1+z)/D_L(z) = 1/D_M(z)$, where
$D_M(z)$ is the proper motion distance.  This distance measure varies
more strongly with $z$ at low redshift than at high redshift.  (See
\cite{h99} for a plot of $D_M(z)$.)  Consequently, when moving from $z
= 1$ to $z = 3$, the amplitude, and thus the SNR, decreases more than
when moving from $z = 3$ to $z = 5$.  For lower mass binaries, this
amplitude decrease plays a bigger role in the loss of SNR than does
redshifting the spectrum to lower frequency; most of the SNR is
accumulated late in the inspiral, where the orbits are in a relatively
flat region of the sensitivity curve.
 
By contrast, for the highest mass binaries, redshifting of the
spectrum can have a dramatic effect.  So much of their signal is moved
out of band that LISA may measure their waves for only a very short
time.  As such, measurement may not provide sufficient information to
constrain 15 parameters.  This is reflected in the high condition
numbers associated with such cases.  Their Fisher matrices are thus
nearly singular, and their inverses are untrustworthy.  In fact,
measurement error in these binaries actually {\em degrades} when
precession is included.  The time in band is too short for precession
effects to accumulate.  They do not aid parameter estimation; instead,
the need to fit extra parameters causes errors to be worse.

\begin{widetext}

\begin{center}
\end{center}

\begin{table}[ht]
\begin{center}
\begin{tabular}{|c|c||c|c|c|c||c|c||c|c|}
\hline
\multirow{2}{*}{$m_1\ (M_\odot)$} & \multirow{2}{*}{$m_2\ (M_\odot)$} & \multirow{2}{*}{$\Delta m_1/m_1$} & \multirow{2}{*}{$\Delta m_2/m_2$} & \multirow{2}{*}{$\Delta \chi_1$} & \multirow{2}{*}{$\Delta \chi_2$} & $\Delta {\mathcal M}/{\mathcal M}$ & $\Delta {\mathcal M}/{\mathcal M}$ & $\Delta \mu/\mu$ & $\Delta \mu/\mu$ \\
& & & & & & (no precession) & (precession) & (no precession) & (precession) \\
\hline \hline
$10^5$ & $10^5$ & 0.000783 & 0.000782 & 0.00415 & 0.00414 & --- & --- & --- & ---  \\ 
\hline
$3\times 10^5$ & $10^5$ & 0.000667 & 0.000541 & 0.00157 & 0.00306 & $5.92 \times 10^{-5}$ & $5.51 \times 10^{-6}$ & 0.0114 & 0.000239  \\ 
\hline
$3\times 10^5$ & $3\times 10^5$ & 0.00109 & 0.00109 & 0.00539 & 0.00536 & --- & --- & --- & ---  \\ 
\hline
$10^6$ & $10^5$ & 0.000629 & 0.000440 & 0.00102 & 0.00440 & 0.000156 & $1.18 \times 10^{-5}$ & 0.0180 & 0.000343  \\ 
\hline
$10^6$ & $3\times 10^5$ & 0.00111 & 0.000882 & 0.00256 & 0.00499 & 0.000170 & $1.19 \times 10^{-5}$ & 0.0274 & 0.000423 \\ 
\hline
$10^6$ & $10^6$ & 0.00195 & 0.00195 & 0.00902 & 0.00897 & --- & --- & --- & ---  \\
\hline
$3\times 10^6$ & $3\times 10^5$ & 0.000988 & 0.000691 & 0.00137 & 0.00563 & 0.000583 & $2.53 \times 10^{-5}$ & 0.0550 & 0.000539 \\ 
\hline
$3\times 10^6$ & $10^6$ & 0.00238 & 0.00192 & 0.00380 & 0.00674 & 0.00117 & $4.19 \times 10^{-5}$ & 0.135 & 0.000849 \\ 
\hline
$3\times 10^6$ & $3\times 10^6$ & 0.00584 & 0.00582 & 0.0271 & 0.0275 & --- & --- & --- & --- \\ 
\hline
$10^7$ & $10^6$ & 0.00239 & 0.00177 & 0.00233 & 0.0122 & 0.00770 & 0.000174 & 0.469 & 0.00140 \\ 
\hline
$10^7$ & $3\times 10^6$ & 0.00814 & 0.00671 & 0.00829 & 0.0159 & 0.00851 & 0.000436 & 0.607 & 0.00332 \\ 
\hline
$10^7$ & $10^7$ & 0.0804 & 0.0802 & 0.492 & 0.493 & --- & --- & --- & --- \\ 
\hline
\end{tabular}
\caption{Median errors in intrinsic quantities for $10^4$ binaries of
various masses at $z = 1$, including comparisons with the ``no
precession'' case where possible.  We have omitted the errors in chirp
mass and reduced mass for equal mass binaries because that
parameterization of the waveform fails the Gaussian approximation at
those points.}
\label{table:intrinsic1}
\end{center}
\end{table}

\begin{table}[ht]
\begin{center}
\begin{tabular}{|c|c||c|c|c|c||c|c||c|c|}
\hline
\multirow{2}{*}{$m_1\ (M_\odot)$} & \multirow{2}{*}{$m_2\ (M_\odot)$} & \multirow{2}{*}{$\Delta m_1/m_1$} & \multirow{2}{*}{$\Delta m_2/m_2$} & \multirow{2}{*}{$\Delta \chi_1$} & \multirow{2}{*}{$\Delta \chi_2$} & $\Delta {\mathcal M}/{\mathcal M}$ & $\Delta {\mathcal M}/{\mathcal M}$ & $\Delta \mu/\mu$ & $\Delta \mu/\mu$ \\
& & & & & & (no precession) & (precession) & (no precession) & (precession) \\
\hline \hline
$10^5$ & $10^5$ & 0.00362 & 0.00362 & 0.0187 & 0.0185 & --- & --- & --- & --- \\
\hline
$3\times 10^5$ & $10^5$ & 0.00363 & 0.00294 & 0.00879 & 0.0171 & 0.000406 & $3.31 \times 10^{-5}$ & 0.0715 & 0.00130 \\ 
\hline
$3\times 10^5$ & $3\times 10^5$ & 0.00569 & 0.00569 & 0.0271 & 0.0269 & --- & --- & --- & --- \\ 
\hline
$10^6$ & $10^5$ & 0.00330 & 0.00231 & 0.00498 & 0.0208 & 0.00120 & $7.09 \times 10^{-5}$ & 0.128 & 0.00180 \\ 
\hline
$10^6$ & $3\times 10^5$ & 0.00648 & 0.00517 & 0.0120 & 0.0229 & 0.00174 & $9.17 \times 10^{-5}$ & 0.228 & 0.00248 \\
\hline
$10^6$ & $10^6$ & 0.0138 & 0.0139 & 0.0627 & 0.0630 & --- & --- & --- & --- \\ 
\hline
$3\times 10^6$ & $3\times 10^5$ & 0.00569 & 0.00402 & 0.00664 & 0.0287 & 0.00633 & 0.000241 & 0.456 & 0.00314 \\ 
\hline
$3\times 10^6$ & $10^6$ & 0.0181 & 0.0148 & 0.0223 & 0.0386 & 0.00708 & 0.000554 & 0.596 & 0.00658 \\ 
\hline
$3\times 10^6$ & $3\times 10^6$ & 0.0744 & 0.0737 & 0.412 & 0.415 & --- & --- & --- & ---  \\ 
\hline
$10^7$ & $10^6$ & 0.0301 & 0.0283 & 0.0256 & 0.177 & 0.0189 & 0.00506 & 0.690 & 0.0231 \\ 
\hline
$10^7$ & $3\times 10^6$ & 0.434 & 0.359 & 0.282 & 0.448 & 0.0182 & 0.0428 & 0.643 & 0.180 \\ 
\hline
$10^7$ & $10^7$ & 12.1 & 12.0 & 62.2 & 61.5 & --- & --- & --- & ---  \\ 
\hline
\end{tabular}
\caption{Median errors in intrinsic quantities for $10^4$ binaries of various masses at $z = 3$.}
\label{table:intrinsic3}
\end{center}
\end{table}

\begin{table}[ht]
\begin{center}
\begin{tabular}{|c|c||c|c|c|c||c|c||c|c|}
\hline
\multirow{2}{*}{$m_1\ (M_\odot)$} & \multirow{2}{*}{$m_2\ (M_\odot)$} & \multirow{2}{*}{$\Delta m_1/m_1$} & \multirow{2}{*}{$\Delta m_2/m_2$} & \multirow{2}{*}{$\Delta \chi_1$} & \multirow{2}{*}{$\Delta \chi_2$} & $\Delta {\mathcal M}/{\mathcal M}$ & $\Delta {\mathcal M}/{\mathcal M}$ & $\Delta \mu/\mu$ & $\Delta \mu/\mu$ \\
& & & & & & (no precession) & (precession) & (no precession) & (precession) \\
\hline \hline
$10^5$ & $10^5$ & 0.00791 & 0.00792 & 0.0392 & 0.0389 & --- & --- & --- & ---  \\ 
\hline
$3\times 10^5$ & $10^5$ & 0.00811 & 0.00658 & 0.0193 & 0.0359 & 0.00103 & $8.00 \times 10^{-5}$ & 0.172 & 0.00290 \\ 
\hline
$3\times 10^5$ & $3\times 10^5$ & 0.0134 & 0.0134 & 0.0615 & 0.0616 & --- & --- & --- & --- \\ 
\hline
$10^6$ & $10^5$ & 0.00718 & 0.00502 & 0.00993 & 0.0409 & 0.00326 & 0.000184 & 0.305 & 0.00391 \\ 
\hline
$10^6$ & $3\times 10^5$ & 0.0156 & 0.0124 & 0.0249 & 0.0460 & 0.00427 & 0.000289 & 0.469 & 0.00596 \\ 
\hline
$10^6$ & $10^6$ & 0.0424 & 0.0423 & 0.197 & 0.200 & --- & --- & --- & --- \\ 
\hline
$3\times 10^6$ & $3\times 10^5$ & 0.0161 & 0.0117 & 0.0158 & 0.0808 & 0.0115 & 0.00103 & 0.643 & 0.00922 \\ 
\hline
$3\times 10^6$ & $10^6$ & 0.0576 & 0.0475 & 0.0606 & 0.107 & 0.0108 & 0.00265 & 0.635 & 0.0214 \\ 
\hline
$3\times 10^6$ & $3\times 10^6$ & 0.396 & 0.391 & 2.43 & 2.44 & --- & --- & --- & --- \\ 
\hline
$10^7$ & $10^6$ & 0.279 & 0.282 & 0.208 & 1.41 & 0.0374 & 0.0640 & 0.704 & 0.232 \\ 
\hline
$10^7$ & $3\times 10^6$ & 10.1 & 8.41 & 6.10 & 7.61 & 0.106 & 1.11 & 0.769 & 4.28 \\ 
\hline
$10^7$ & $10^7$ & 2280 & 2290 & 10300 & 9900 & --- & --- & --- & --- \\ 
\hline
\end{tabular}
\caption{Median errors in intrinsic quantities for $10^4$ binaries of
various masses at $z = 5$.  The results for the highest masses are
meaningless --- the parameters are completely undetermined.}
\label{table:intrinsic5}
\end{center}
\end{table}
\end{widetext}

\subsection{Sky position and distance to source}
\label{sec:resII}

We now focus on extrinsic parameters, the sky position and the
luminosity distance to the source.  We find that the determination of
these parameters is likewise improved when precession physics is taken
into account, though not as strongly as for intrinsic parameters.
This might be expected, since precession is an intrinsic effect local
to the binary and has no direct dependence on these extrinsic
parameters.  Precession's impact on the extrinsic parameters is
somewhat more indirect --- it largely improves their determination by
reducing the (otherwise quite strong) correlation between sky position
and the orbital angular momentum direction $\mathbf{\hat{L}}$ and
between these angles and the source's luminosity distance.

In our analysis, a binary's position on the sky is characterized by
the two parameters $\boldsymbol{\theta} = (\bar{\mu}_N = \cos
\bar{\theta}_N, \bar{\phi}_N)$.  Consider the subspace containing just
these two parameters.  The Fisher matrix is a $2 \times 2$ matrix
which characterizes the probability density for the true parameters
given a measured signal [see \eqref{eq:linearizedprob}]:

\begin{equation}
p(\boldsymbol{\tilde{\theta}}|s) \propto \exp
\left(-\frac{1}{2}\Gamma_{ab}\delta \theta^a \delta \theta^b \right)
\, .
\label{eq:undiagonalized}
\end{equation}
To accurately describe the error ellipse on the sky, we need to
manipulate the right hand side of \eqref{eq:undiagonalized} in several
ways.  First, we change coordinates from $\bar{\mu}_N$ to
$\bar{\theta}_N$, while leaving $\bar{\phi}_N$ alone; schematically,
we can represent this as $\theta^a \rightarrow \theta^{a^\prime}$.
This transformation gives $\delta \bar{\theta}_N =
(d\bar{\theta}_N/d\bar{\mu}_N)\delta \bar{\mu}_N = -\delta
\bar{\mu}_N/ \sin \bar{\theta}_N $.

\begin{figure}[htb]
\includegraphics[scale=0.54]{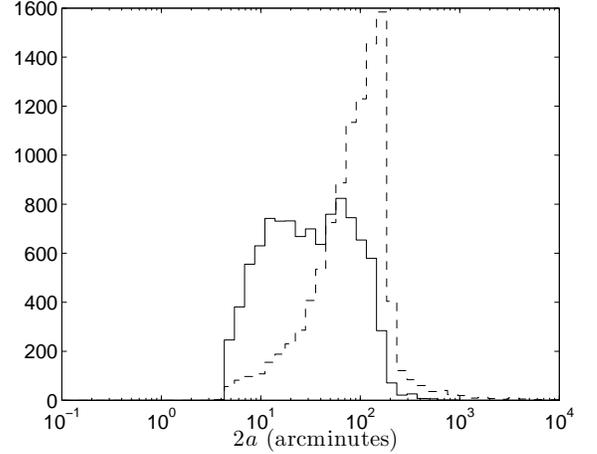}
\caption{Distribution of the major axis of the sky position error
ellipse, $2a$, for $10^4$ binaries with $m_1 = 10^6 M_\odot$ and $m_2
= 3 \times 10^5 M_\odot$ at $z = 1$.  The dashed line is the
precession-free calculation; the solid line includes precession.  Sky
position, as an extrinsic parameter, is improved somewhat indirectly
by precession; therefore, the improvement is less than for the
masses.}
\label{fig:2acomparison}
\end{figure}

\begin{figure}[htb]
\includegraphics[scale=0.54]{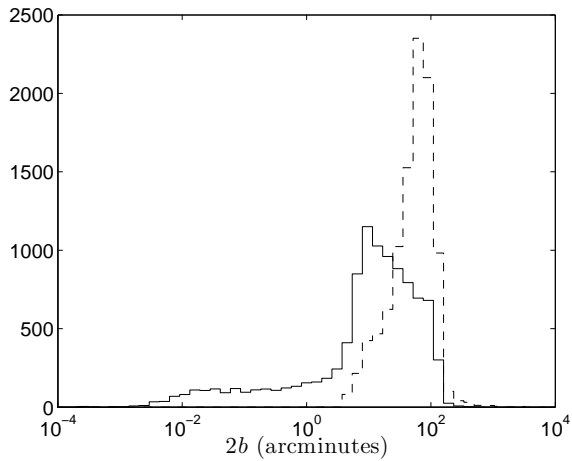}
\caption{Distribution of the minor axis of the sky position error
ellipse, $2b$, for $10^4$ binaries with $m_1 = 10^6 M_\odot$ and $m_2
= 3 \times 10^5 M_\odot$ at $z = 1$.  The dashed line is the
precession-free calculation; the solid line includes precession.}
\label{fig:2bcomparison}
\end{figure}

Next, we need to redefine what we mean by ``error'' in order to make
the results more relevant to observations.  To do so, we define the
``proper'' angular errors $\delta \bar{\theta}_N^p = \delta
\bar{\theta}_N$ and $\delta \bar{\phi}_N^p = \sin \bar{\theta}_N\delta
\bar{\phi}_N$.  The proper angular errors are just the normal
coordinate errors rescaled by the metric of the sphere to correctly
account for the proper size of a segment $\delta \bar{\phi}_N$ at
different $\bar{\theta}_N$.  Substituting all of these changes into
\eqref{eq:undiagonalized}, we obtain

\begin{equation}
p(\boldsymbol{\tilde{\theta}}|s) \propto \exp
\left(-\frac{1}{2}\Gamma^p_{a^\prime b^\prime}\delta
\theta^{a^\prime}_p \delta \theta^{b^\prime}_p \right) \, ,
\label{eq:properundiagonalized}
\end{equation}
where we have defined a proper Fisher matrix for the parameters
$(\bar{\theta}_N, \bar{\phi}_N)$.  In terms of the original Fisher
matrix, the elements are

\begin{align}
\Gamma^p_{\bar{\theta}_N \bar{\theta}_N} &= \sin^2 \bar{\theta}_N
\Gamma_{\bar{\mu}_N \bar{\mu}_N} \, , \\
\Gamma^p_{\bar{\theta}_N \bar{\phi}_N} &= \Gamma^p_{\bar{\phi}_N
\bar{\theta}_N} = -\Gamma_{\bar{\mu}_N \bar{\phi}_N} \, , \\
\Gamma^p_{\bar{\phi}_N \bar{\phi}_N} &= \csc^2 \bar{\theta}_N
\Gamma_{\bar{\phi}_N \bar{\phi}_N} \, .
\end{align}
Finally, we diagonalize the Fisher matrix by rotating our
parameterization, $\theta^{a^\prime} \rightarrow \theta^{\hat{a}}$,
such that the probability \eqref{eq:properundiagonalized} becomes

\begin{equation}
p(\boldsymbol{\tilde{\theta}}|s) \propto \exp \left[-\left(\frac{(\delta
\theta^{\hat{1}}_p)^2}{2(\sigma^p_{\hat{1}})^2} + \frac{(\delta
\theta^{\hat{2}}_p)^2}{2(\sigma^p_{\hat{2}})^2}\right)\right] \, .
\label{eq:diagonalized}
\end{equation}
In these coordinates, the covariance matrix is
\begin{equation}
\Sigma^{\hat{a}\hat{b}}_p = \left[
\begin{array}{cc}
(\sigma^p_{\hat{1}})^2 & 0 \\
0 & (\sigma^p_{\hat{2}})^2 
\end{array}
\right] \, .
\label{eq:diagonalcovariance}
\end{equation}
Following Cutler \cite{c98}, we define the error ellipse such that the
probability that the source lies outside the error ellipse is
$e^{-1}$.  The semiaxes of the error ellipse are given by
$\sqrt{2(\sigma^p_{\hat{1},\hat{2}})^2}$.  These quantities follow
from the eigenvalues of the covariance matrix
\eqref{eq:diagonalcovariance}; since eigenvalues are invariant under
rotation, we can calculate them before performing the rotation.  In
terms of our original covariance matrix, the major axis $2a$ and minor
axis $2b$ of the ellipse are given by

\begin{equation}
\begin{split}
& 2\left[
\vphantom{\sqrt{(\csc^2\bar{\theta}_N\Sigma^{\bar{\mu}_N\bar{\mu}_N} -
\sin^2\bar{\theta}_N\Sigma^{\bar{\phi}_N\bar{\phi}_N})^2 +
4(\Sigma^{\bar{\mu}_N\bar{\phi}_N})^2}}
\csc^2\bar{\theta}_N\Sigma^{\bar{\mu}_N\bar{\mu}_N} +
\sin^2\bar{\theta}_N\Sigma^{\bar{\phi}_N\bar{\phi}_N} \pm \right. \\
& \left. 
\sqrt{(\csc^2\bar{\theta}_N\Sigma^{\bar{\mu}_N\bar{\mu}_N} -
\sin^2\bar{\theta}_N\Sigma^{\bar{\phi}_N\bar{\phi}_N})^2 +
4(\Sigma^{\bar{\mu}_N\bar{\phi}_N})^2}
\right]^{1/2} \, , 
\end{split} 
\label{eq:2a2b}
\end{equation}
taking the plus and minus for major and minor axes, respectively.  We
also find the area of the error ellipse:

\begin{equation}
\Delta \Omega_N = \pi ab =
2\pi\sqrt{\Sigma^{\bar{\mu}_N\bar{\mu}_N}\Sigma^{\bar{\phi}_N\bar{\phi}_N}
- (\Sigma^{\bar{\mu}_N\bar{\phi}_N})^2} \, .
\label{eq:deltaOmS}
\end{equation}
Many previous analyses have reported the ellipse's area $\Delta
\Omega_N$ or $\sqrt{\Delta\Omega_N}$, the side of a square of
equivalent area, as the sky position error {\cite{c98,bbw05,hh05}}.
Information about the ellipse's shape, crucial input to coordinating
GW observations with telescopes, is not included in such a measure.
By examining both $2a$ and $2b$, this information is restored.  Figure
\ref{fig:2acomparison} shows the major axis of the error ellipse $2a$
for both the original code, with no precession, and the code including
precession effects.  Figure \ref{fig:2bcomparison} shows the same for
the minor axis $2b$.  (Note that these figures cannot tell us which
major axis is associated with which minor axis; that information is
lost in the construction of the histograms.)

Compared to the code which does include precession physics, the median
of both distributions is reduced by about half an order of magnitude.
The minor axis distribution also shows a long tail of very small
errors.  In those cases, the position would be very well-constrained
in one direction.

\begin{figure}[htb]
\includegraphics[scale=0.54]{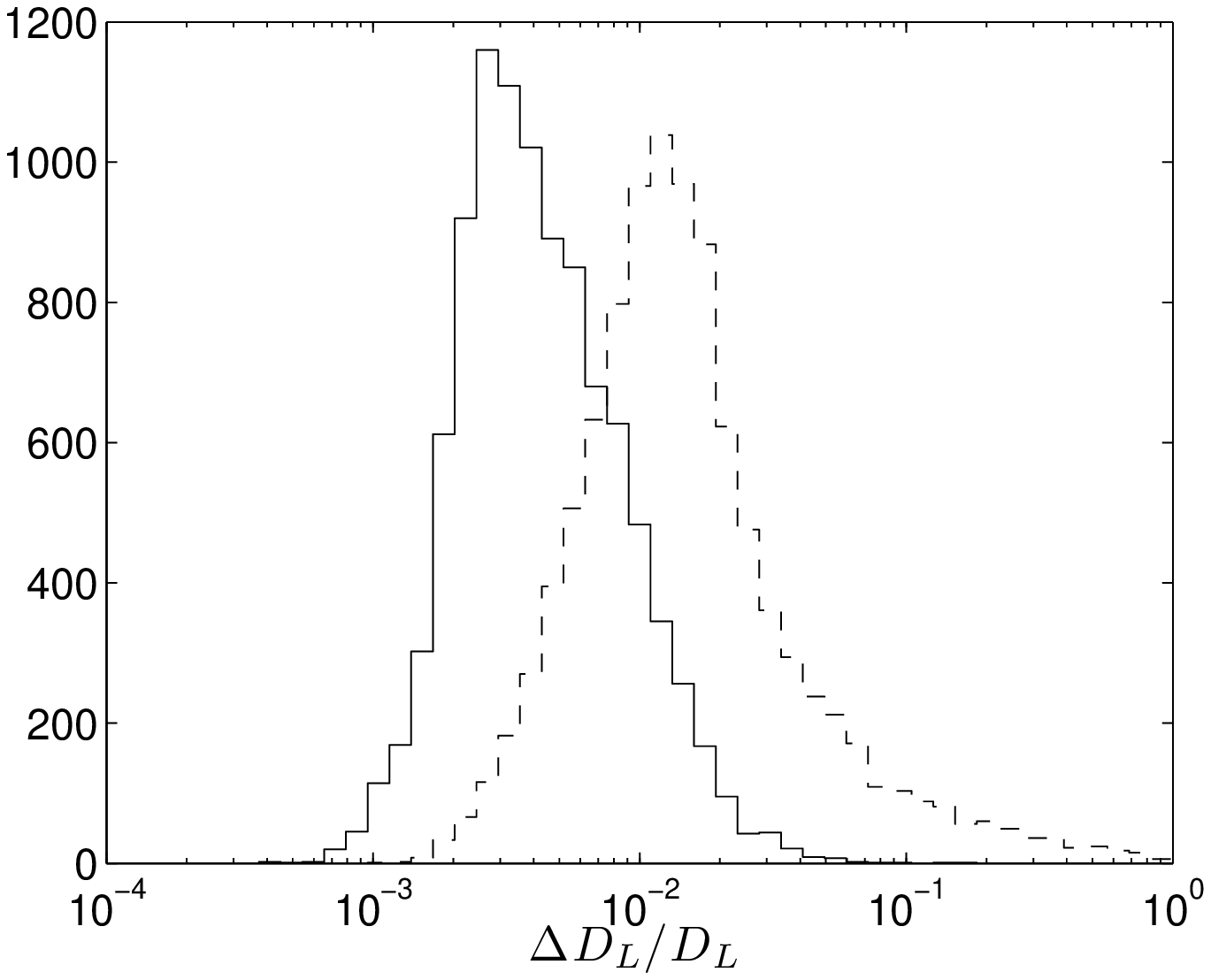}
\caption{Distribution of errors in the luminosity distance for $10^4$
binaries with $m_1 = 10^6 M_\odot$ and $m_2 = 3 \times 10^5 M_\odot$
at $z = 1$.  The dashed line is the precession-free calculation; the
solid line includes precession.}
\label{fig:Dcomparison}
\end{figure}

Finally, we examine how well distance to the binary is determined.
Figure \ref{fig:Dcomparison} compares $\Delta D_L/D_L$ both with and
without precession physics taken into account.  For this case, the
distance error improves by about a factor of 3.

Table \ref{table:extrinsic1} shows the median extrinsic errors for
binaries of different mass.  For comparison purposes, we include
results that neglect spin precession.  Binaries with the best
determined parameters at this redshift have total mass
$\mbox{several}\times 10^5\,M_\odot \lesssim M_{\rm tot} \lesssim
\mbox{several}\times 10^6\,M_\odot$ --- smaller binaries are not quite
determined so well due to the weakness of their signal, while larger
ones are not determined so well because they radiate fewer cycles in
band.  We also see again that unequal mass binaries give better
results than equal mass binaries due to the impact of mass ratio on
precession effects.  Overall, we find that the major axis of the error
ellipse is on the order of $\mbox{a few} \times 10$ arcminutes, while
the minor axis is a factor of $2-4$ smaller.  This represents an
improvement over the ``no precession'' case by a factor $\sim 2 - 7$
for the major axis and a factor $\sim 2 - 10$ for the minor axis.  The
distance errors are on the order of $0.2\% - 0.7\%$ for most masses, a
factor of $\sim 2 - 7$ improvement.

Tables \ref{table:extrinsic3} and \ref{table:extrinsic5} show the same
results for higher redshift.  We see the same trends as at $z = 1$,
but with some degradation in numerical value.  The sky position errors
reach a few degrees in the major axis and several tens of arcminutes
up to a degree or two in the minor axis.  The distance errors are on
the order of 1 to several percent for most masses.  At the highest
masses, we again see that these parameters are essentially
undetermined and that precession makes things worse by requiring extra
parameters to be fit.

\newpage

\begin{widetext}

\begin{center}
\end{center}

\begin{table}
\begin{center}
\begin{tabular}{|c|c||c|c||c|c||c|c|}
\hline
\multirow{2}{*}{$m_1\ (M_\odot)$} & \multirow{2}{*}{$m_2\ (M_\odot)$} & $2a\ (\mathrm{arcmin})$ & $2a\ (\mathrm{arcmin})$ & $2b\ (\mathrm{arcmin})$ & $2b\ (\mathrm{arcmin})$ & $\Delta D_L/D_L$ & $\Delta D_L/D_L$ \\
& & (no precession) & (precession) & (no precession) & (precession) & (no precession) & (precession) \\
\hline \hline
$10^5$ & $10^5$ & 133 & 27.3 & 84.7 & 13.3 & 0.0193 & 0.00398 \\ 
\hline
$3\times 10^5$ & $10^5$ & 115 & 16.9 & 72.6 & 7.33 & 0.0165 & 0.00240 \\ 
\hline
$3\times 10^5$ & $3\times 10^5$ & 101 & 23.3 & 62.8 & 11.8 & 0.0143 & 0.00357 \\ 
\hline
$10^6$ & $10^5$ & 105 & 27.2 & 65.1 & 6.62 & 0.0149 & 0.00320 \\ 
\hline
$10^6$ & $3\times 10^5$ & 93.1 & 31.3 & 57.5 & 13.2 & 0.0132 & 0.00393 \\ 
\hline
$10^6$ & $10^6$ & 90.1 & 40.2 & 54.1 & 21.9 & 0.0125 & 0.00560 \\
\hline
$3\times 10^6$ & $3\times 10^5$ & 95.0 & 34.1 & 57.3 & 9.20 & 0.0135 & 0.00376 \\ 
\hline
$3\times 10^6$ & $10^6$ & 102 & 32.3 & 56.0 & 14.7 & 0.0135 & 0.00419 \\ 
\hline
$3\times 10^6$ & $3\times 10^6$ & 135 & 43.3 & 68.5 & 22.3 & 0.0182 & 0.00689 \\ 
\hline
$10^7$ & $10^6$ & 149 & 37.6 & 75.2 & 12.2 & 0.0200 & 0.00457 \\ 
\hline
$10^7$ & $3\times 10^6$ & 238 & 42.1 & 119 & 19.0 & 0.0322 & 0.00610\\ 
\hline
$10^7$ & $10^7$ & 466 & 81.3 & 232 & 38.6 & 0.0636 & 0.0250 \\ 
\hline
\end{tabular}
\caption{Median errors in extrinsic quantities for $10^4$ binaries of
various masses at $z = 1$, including comparisons with the ``no
precession'' case.  Note that the given major axis and minor axis are
the medians for each data set and do not correspond to the same
binary.  However, they still represent an average sky position error
ellipse in the following sense: $\sqrt{\pi ab}$, calculated using the
median values of $2a$ and $2b$, differs in most cases by less than $10\%$ 
from the median value of $\sqrt{\Delta \Omega_N}$ calculated
from the covariance matrix and \eqref{eq:deltaOmS} (except at more
extreme mass ratios --- when $m_1/m_2 = 10$, the difference can be
$25\%$).}
\label{table:extrinsic1}
\end{center}
\end{table}

\begin{table}
\begin{center}
\begin{tabular}{|c|c||c|c||c|c||c|c|}
\hline
\multirow{2}{*}{$m_1\ (M_\odot)$} & \multirow{2}{*}{$m_2\ (M_\odot)$} & $2a\ (\mathrm{arcmin})$ & $2a\ (\mathrm{arcmin})$ & $2b\ (\mathrm{arcmin})$ & $2b\ (\mathrm{arcmin})$ & $\Delta D_L/D_L$ & $\Delta D_L/D_L$ \\
& & (no precession) & (precession) & (no precession) & (precession) & (no precession) & (precession) \\
\hline \hline
$10^5$ & $10^5$ & 432 & 81.0 & 271 & 40.8 & 0.0617 & 0.0123 \\ 
\hline
$3\times 10^5$ & $10^5$ & 389 & 92.5 & 242 & 39.5 & 0.0551 & 0.0126 \\ 
\hline
$3\times 10^5$ & $3\times 10^5$ & 356 & 142 & 220 & 75.7 & 0.0502 & 0.0201 \\ 
\hline
$10^6$ & $10^5$ & 379 & 141 & 233 & 36.6 & 0.0550 & 0.0155 \\ 
\hline
$10^6$ & $3\times 10^5$ & 359 & 129 & 215 & 56.7 & 0.0500 & 0.0161 \\ 
\hline
$10^6$ & $10^6$ & 416 & 158 & 224 & 84.3 & 0.0556 & 0.0237 \\ 
\hline
$3\times 10^6$ & $3\times 10^5$ & 425 & 132 & 233 & 40.3 & 0.0568 & 0.0153 \\ 
\hline
$3\times 10^6$ & $10^6$ & 599 & 142 & 302 & 64.6 & 0.0809 & 0.0193 \\ 
\hline
$3\times 10^6$ & $3\times 10^6$ & 990 & 224 & 494 & 111 & 0.134 & 0.0422 \\ 
\hline
$10^7$ & $10^6$ & 1320 & 206 & 648 & 78.5 & 0.178 & 0.0293 \\ 
\hline
$10^7$ & $3\times 10^6$ & 2380 & 297 & 1180 & 152 & 0.326 & 0.0805 \\ 
\hline
$10^7$ & $10^7$ & 6820 & 2000 & 3390 & 583 & 0.935 & 2.41 \\ 
\hline
\end{tabular}
\caption{Median errors in extrinsic quantities for $10^4$ binaries of
various masses at $z = 3$.}
\label{table:extrinsic3}
\end{center}
\end{table}

\begin{table}
\begin{center}
\begin{tabular}{|c|c||c|c||c|c||c|c|}
\hline
\multirow{2}{*}{$m_1\ (M_\odot)$} & \multirow{2}{*}{$m_2\ (M_\odot)$} & $2a\ (\mathrm{arcmin})$ & $2a\ (\mathrm{arcmin})$ & $2b\ (\mathrm{arcmin})$ & $2b\ (\mathrm{arcmin})$ & $\Delta D_L/D_L$ & $\Delta D_L/D_L$ \\
& & (no precession) & (precession) & (no precession) & (precession) & (no precession) & (precession) \\
\hline \hline
$10^5$ & $10^5$ & 729 & 169 & 456 & 85.7 & 0.104 & 0.0260 \\ 
\hline
$3\times 10^5$ & $10^5$ & 676 & 217 & 419 & 95.8 & 0.0957 & 0.0284 \\ 
\hline
$3\times 10^5$ & $3\times 10^5$ & 650 & 295 & 395 & 161 & 0.0917 & 0.0409 \\ 
\hline
$10^6$ & $10^5$ & 686 & 248 & 416 & 66.8 & 0.0983 & 0.0273 \\ 
\hline
$10^6$ & $3\times 10^5$ & 716 & 233 & 404 & 101 & 0.0961 & 0.0294 \\ 
\hline
$10^6$ & $10^6$ & 976 & 315 & 497 & 162 & 0.132 & 0.0501 \\ 
\hline
$3\times 10^6$ & $3\times 10^5$ & 986 & 265 & 507 & 86.4 & 0.133 & 0.0318 \\ 
\hline
$3\times 10^6$ & $10^6$ & 1620 & 304 & 810 & 139 & 0.220 & 0.0436 \\ 
\hline
$3\times 10^6$ & $3\times 10^6$ & 2930 & 538 & 1460 & 260 & 0.400 & 0.140 \\ 
\hline
$10^7$ & $10^6$ & 5080 & 577 & 2480 & 290 & 0.689 & 0.124 \\ 
\hline
$10^7$ & $3\times 10^6$ & 10500 & 1720 & 5130 & 621 & 1.42 & 1.24 \\ 
\hline
$10^7$ & $10^7$ & 75500 & 180000 & 35000 & 29600 & 10.3 & 377 \\ 
\hline
\end{tabular}
\caption{Median errors in extrinsic quantities for $10^4$ binaries of
various masses at $z = 5$.  Again, the results for the highest masses
are essentially meaningless--the parameters are completely
undetermined.}
\label{table:extrinsic5}
\end{center}
\end{table}
\end{widetext}

\section{Summary and conclusions}
\label{sec:disc}

The general relativistic precession of black holes in binary systems
can have a strong influence on the binary's dynamics
{\cite{acst94,s04}} and thus upon the GWs that it generates.  It has
been known for some time that it will be necessary to take these
dynamics into account in order to detect these black holes in noisy
detector data {\cite{bcv03,bcv041,bcv042,bcv05,gkv03,gk03,gikb04}};
clearly, taking these dynamics into account will be just as (if not
more) important for the complementary problem of determining the
parameters which characterize a detected system.  Vecchio {\cite{v04}}
first demonstrated that, by taking into account precession physics,
quite a few near degeneracies among binary source parameters can be
broken, making our estimates for how accurately they can be determined
more optimistic.  This analysis largely confirms and extends Vecchio's
pioneering work.  By taking the equations of motion to higher order to
include spin-spin couplings, and by surveying measurement accuracy as
a function of mass ratio, we have found that the improvement noted by
Vecchio holds rather broadly.  The degeneracy breaking due to
precession physics is a rather robust phenomenon.

Two conclusions from this work are particularly important with regard
to the astrophysical reach of future LISA measurements.  The first is
that modeling spin-precession physics makes it possible to determine
the magnitudes of the spins of the black holes which constitute the
binary.  If the spins are rapid, they can be measured quite accurately
(as good as $0.1 \%$ accuracy for high spin, low redshift systems) due
to the strong modulation imposed on the signal by their interaction.
Coupled with the fact that the black hole masses can likewise be
measured with good precision, this suggests that LISA will be a
valuable tool for tracking the evolution of both mass and spin over
cosmic time.  Such observations could provide a direct window into the
growth of cosmological structures.  Measuring spin may also make it
possible to indirectly test the black hole area theorem {\cite{h71}}.
The requirement that black hole area can only grow implies a
consistency relation between the initial and final masses and spins.
By measuring the initial masses and spins through the inspiral, and
the mass and spin of the merged remnant hole through the ringdown
waves {\cite{dkkfgl04,bcw06}}, we can check this consistency relation
in a manner analogous to the mass loss test proposed in {\cite{hm05}}.
We intend to investigate whether this test is feasible in future work.

Second, we confirm Vecchio's result that precession breaks
degeneracies between the angles which determine a binary's orientation
and its position on the sky, improving the accuracy with which sky
position can be fixed using GWs alone.  At low redshift ($z \sim 1$),
we find that sources can be localized to within an ellipse whose major
axis is typically $\mbox{a few}\times 10$ arcminutes across and whose
minor axis is typically a factor $\sim 2-4$ smaller.  This is small
enough that searching the GW pixel for an electromagnetic counterpart
to the merger event should not be too arduous a task \cite{kfhm06}.
For mergers at higher redshift, the waves weaken and the source is not
so well localized.  The field which would need to be searched for
sources at $z \sim 3 - 5$ is typically a few degrees across in the
long axis and tens of arcminutes to a degree or two in the short
direction --- a rather more difficult challenge, but not hopeless.  We
intend to more thoroughly investigate the nature of localization with
spin precession, including how the pixel evolves with observation time
up to final merger, in future work.

As mentioned in the Introduction, our analysis makes many assumptions
and approximations which are likely to affect our results; a goal of
future work will be to lift these approximations.  One major concern
is the Gaussian approximation we have taken to the likelihood
function.  As already discussed, this approximation is known to be
good when the SNR is ``large'' {\cite{f92,cf94}}; however, it is not
apparent what large really means, particularly given that we are
fitting for 15 parameters.  Lifting this simplifying approximation can
be done by simply computing the likelihood function
(\ref{eq:signalprob}) directly and examining how well parameters are
thereby determined.  In the context of GW measurements, Markov
Chain-Monte Carlo techniques have been investigated and found to be
very useful {\cite{cm01,cdwm04,umdvwc04}}; the application of these
techniques to LISA measurement problems is now being rather actively
investigated {\cite{cc05,cp06_1,cp06_2}}.

Because we have used the Gaussian approximation (among other
simplifications taken in this analysis), we cannot claim that this
analysis gives a definitive statement about the accuracy with which
LISA could measure binary black hole source parameters.  However, it
is certainly {\it indicative} of the accuracy which we expect LISA to
achieve.  In particular, we are confident that the trends we have seen
as parameters are varied (e.g., masses, redshift, spin magnitude) are
robust.  Most importantly, it is very clear that the influence of
spin-induced precession upon the measured waveform allows parameters
to be measured to greater accuracy than before.\\

\acknowledgments

This work benefited greatly from discussions with and feedback from
members of the LISA International Science Team, especially Neil
Cornish, E.\ Sterl Phinney, and Thomas Prince. We also thank Bence
Kocsis for very useful discussions about parameterizing the sky
position error ellipse, Shane Larson for providing code to process his
noise files, and Jeremy Schnittman for helpful comments about the
precession equations. We are especially grateful to Emanuele Berti for
discussions regarding integrators, which led to a drastic improvement
in the ability of our code to perform large Monte Carlo surveys, as
well as many insightful comments about this manuscript.  This work was
supported by NASA Grant NAGW-12906 and NSF Grant PHY-0449884.  S.A.H.
gratefully acknowledges the support of the MIT Class of 1956 Career
Development fund.

\end{document}